\def\etal{{et~al.\thinspace}}
\def\mearth{{\rm\,M_\oplus}}

\def\bbc{{\beta/\beta_{crit}}}

\documentclass[apjfonts]{emulateapj}
\slugcomment{ApJ, in press}

\begin{document}

\shorttitle{Planet-planet scattering in planetesimal disks}
\shortauthors{Raymond, Armitage \& Gorelick}

\title{Planet-planet scattering in planetesimal disks II: \\
Predictions for outer extrasolar planetary systems}

\author{Sean N. Raymond\altaffilmark{1,2,3}, 
Philip J. Armitage\altaffilmark{4,5},
\& Noel Gorelick\altaffilmark{6}}

\altaffiltext{1}{Universit{\'e} de Bordeaux, Observatoire Aquitain des Sciences de l'Univers, 2 rue de l'Observatoire, BP 89, F-33271 Floirac Cedex, France}
\altaffiltext{2}{CNRS, UMR 5804, Laboratoire d'Astrophysique de Bordeaux, 2 rue de l'Observatoire, BP 89, F-33271 Floirac Cedex, France}
\altaffiltext{3}{Center for Astrophysics and Space Astronomy, 389 UCB,
University of Colorado, Boulder CO 80309, USA}
\altaffiltext{4}{JILA, University of Colorado, Boulder CO 80309, USA; pja@jilau1.colorado.edu}
\altaffiltext{5}{Department of Astrophysical and Planetary Sciences, University of Colorado, Boulder CO 80309}
\altaffiltext{6}{Google, Inc., 1600 Amphitheatre Parkway, Mountain View, CA 94043, USA}

\begin{abstract}
We develop an idealized dynamical model to predict the typical properties
of outer extrasolar planetary systems, at radii comparable to the Jupiter
to Neptune region of the Solar System. The model is based upon the
hypothesis that dynamical evolution in outer planetary systems is
controlled by a combination of planet-planet scattering and planetary
interactions with an exterior disk of small bodies (``planetesimals"). Our
results are based on 5,000 long duration N-body simulations that follow the
evolution of three planets from a few to 10 AU, together with a
planetesimal disk containing 50~$\mearth$ from 10-20 AU. For large planet
masses ($M \gtrsim M_{\rm Sat}$) the model recovers the observed
eccentricity distribution of extrasolar planets. For lower mass planets the
range of outcomes in models with disks is far greater than is seen in
isolated planet-planet scattering. Common outcomes include strong
scattering among massive planets, sudden jumps in eccentricity due to
resonance crossings driven by divergent migration, and re-circularization
of scattered low-mass planets in the outer disk. We present distributions
of the eccentricity and inclination that result, and discuss how they vary
with planet mass and initial system architecture.  In agreement with other
studies, we find that the currently observed eccentricity distribution
(derived primarily from planets at $a \lesssim 3$~AU) is consistent with
isolated planet-planet scattering. We explain the observed mass dependence
-- which is in the opposite sense from that predicted by the simplest
scattering models -- as a consequence of strong correlations between planet
masses in the same system. At somewhat larger radii initial planetary mass
correlations and disk effects can yield similar modest changes to the
eccentricity distribution. Nonetheless, {\em strong} damping of
eccentricity for low mass planets at large radii appears to be a secure
signature of the dynamical influence of disks. Radial velocity measurements 
capable of detecting planets with $K \approx 5 \ {\rm m \ s}^{-1}$ and periods 
in excess of 10~years will provide constraints on this regime. Finally, we present an
analysis of the predicted separation of planets in two planet systems, and
of the population of planets in mean motion resonances (MMRs). We show
that, if there are systems with $\sim$Jupiter-mass planets that avoid close
encounters, the planetesimal disk acts as a damping mechanism and populates
mean motion resonances (MMRs) at a very high rate (50-80\%).  In many
cases, resonant chains (in particular the 4:2:1 Laplace resonance) are set
up among all three planets.  We expect such resonant chains to be common
among massive planets in outer planetary systems.
\end{abstract}


\section{Introduction}
Observations of planetary systems containing massive planets suggest that
the final architecture of planetary systems is often determined by
evolutionary processes that occur {\em after} the planets have formed and
accreted much of their mass. In the Solar System strong evidence for
early evolution comes from the orbits of small bodies in the Kuiper Belt
(Jewitt \& Luu 1993), many of which occupy mean-motion resonances (MMRs)
with Neptune (Chiang \etal 2007). The only known explanation for this
unusual distribution of orbital properties involves the resonant capture
(Goldreich 1965) of Pluto and other Kuiper Belt Objects (KBOs) by Neptune
in the course of a slow expansion of its orbit during the early history of
the Solar System (Malhotra 1993, 1995). In extrasolar planetary systems
inward orbital migration is required in order to explain
``hot Jupiters", massive extrasolar planets with orbital radii $a
\lesssim 0.1~{\rm AU}$ (Mayor \& Queloz 1995), whose existence is inconsistent 
with {\em in situ} giant planet formation (Bodenheimer \etal 2000). Even
more striking is the fact that extrasolar planets with $a \gtrsim 0.1~{\rm
AU}$ display a broad distribution of eccentricities (Marcy \etal 2005) that
is at odds with the expectation that planets form in near-circular
orbits. In addition to these general properties, the unusual
characteristics of individual systems such as XO-3, whose orbital plane
does not coincide with the stellar equator (H{\'e}brard \etal 2008; Winn
\etal 2009), and the existence of resonant multiple planet systems such as
GJ~876 (Marcy \etal 2001), demand explanation in terms of post-formation
dynamical evolution.

The action of three physical mechanisms: gas disk migration, planetesimal
disk migration, and planet-planet scattering, can lead to large-scale
changes in planetary orbital elements (the semi-major axis $a$,
eccentricity $e$, and inclination $i$). The basic physics underlying each
of these processes is now moderately well-understood (for 
reviews, see e.g. Papaloizou \& Terquem 2006; Armitage 2010). Gas disk migration is mediated by
the exchange of energy and angular momentum between a planet and the gas
disk at Lindblad resonances and in the co-orbital region (Goldreich \& Tremaine 1980).
In the regime relevant to massive planet migration the coupling is strong,
and the end result is that the planet's semi-major axis changes on a time
scale given to order of magnitude by the viscous time scale of the
protoplanetary disk (Lin \& Papaloizou 1986). Modest eccentricity growth
may accompany decay of the semi-major axis (Ogilvie \& Lubow 2003;
Goldreich \& Sari 2003; D'Angelo \etal 2006; Moorhead \& Adams 2008), while
any mutual inclination between planet and disk is damped (Lubow \& Ogilvie 
2001)\footnote{Damping is predicted for a single planet migrating through a
gas disk. If two planets are simultaneously migrating resonant
perturbations {\em between} the planets can overcome the damping and lead
to an increase in $i$ (Thommes \& Lissauer 2003; Lee \& Thommes 2009).}. Planetesimal disk
migration occurs due to the scattering or ejection of a collisionless
population of small bodies\footnote{For convenience we dub these small
bodies ``planetesimals", although they could be substantially smaller or
larger than the 10-100~km planetesimals conventionally envisaged as the
first stage of planet formation. The actual physical size is of little
import provided that the bodies are neither so small that they behave as a
collisional fluid, nor so large that {\em individual} scattering events
significantly perturb the orbit of a planet.}  by a planet (Fernandez \& Ip
1984; Hahn \& Malhotra 1999).  The sense of migration can be inward (if the
planet predominantly ejects planetesimals) or outward (if the planet
scatters bodies from an exterior disk onto lower angular momentum orbits),
and occurs at a rate that depends upon the surface density of the
planetesimal disk (Ida \etal 2000; Gomes \etal 2004; Kirsh \etal 2009). Planetary
eccentricity is damped (Murray \etal 2002). Finally planet-planet
scattering occurs when an initially unstable system of $N$ planets relaxes
under the action of purely gravitational forces. Typically such a system
evolves chaotically up to the development of orbit crossing, which results
in the ejection of (or collisions between) some of the planets. The
survivors have a broad distribution of $e$, non-zero $i$, and have moved
closer to the star as a consequence of the loss of energy (Rasio
\& Ford 1996; Weidenschilling \& Marzari 1996; Lin \& Ida 1997).

Direct observational evidence for the importance of these processes is
limited. For the Solar System the structure of the Kuiper Belt (Chiang
\etal 2003; Levison \& Morbidelli 2003; Hahn \& Malhotra 2005; 
Murray-Clay \& Chiang 2005; Levison \etal 2008) and of the main
asteroid belt (Minton \& Malhotra 2009) is broadly consistent with
predictions based on planetesimal disk-driven migration.
Planetesimal-driven planetary migration can also provide plausible
explanations for otherwise puzzling features of the Solar System such as
the large inclinations of Jupiter's Trojan asteroids and the composition (and 
possibly the timing) of the
Late Heavy Bombardment on the Moon (Tera \etal 1974; Strom \etal 2005; 
Tsiganis \etal 2005; Morbidelli
\etal 2005; Gomes \etal 2005). Taken together the evidence clearly suggests
a dominant dynamical role for planetesimal scattering in the early history
of the outer Solar System. For extrasolar planetary systems, on the other
hand, even the best evidence is circumstantial. It is now well-established
that the eccentricity distribution of massive extrasolar planets matches
the predictions of planet-planet scattering models (Ford \etal 2003; Adams
\& Laughlin 2003; Chatterjee \etal 2008; Juric \& Tremaine 2008; Raymond
\etal 2008a; Thommes \etal 2008a). This does not rule out the possibility that significant
eccentricity excitation occurred during gas disk migration, but it
supports the contention that the progenitors to today's observed systems
were dynamically unstable. The relative role of gas disk migration and
planet-planet scattering in setting up the observed radial distribution of
extrasolar planets is harder to determine.  If planet-planet scattering
occurred in most systems the accompanying change in $a$ of the innermost
planet could have populated much of the observed extrasolar planet region
from a parent population beyond the snowline, especially if $N$ were
moderately large (Papaloizou \& Terquem 2001).  More commonly, however, it
is assumed that the present semi-major axes of not just the hot Jupiters
(Lin \etal 1996), but also the bulk of known extrasolar planets, are the
result of gas disk migration (Trilling \etal 1998).  Models of planetary
migration within evolving disks can be constructed that match the observed
distribution of planets (Armitage \etal 2002; Armitage 2007), but theoretical knowledge of
disks falls some way short of allowing an unambiguous determination of the
importance of migration to be made.

Existing dynamical studies of extrasolar planets have largely focused on
explaining the properties of planets observed at small radii ($a \lesssim 5
\ {\rm AU}$), where gas disks may be important but the mass in small bodies
is assuredly negligible\footnote{An exception is the work of (Murray \etal
1998), who considered inward migration of massive planets within an
extremely massive planetesimal disk.}. Our goal in this paper is to study
the dynamics of extrasolar planetary systems at larger orbital radii, where
dynamical effects associated with planetesimal disks become important.  We
aim to map out the final architecture of planetary systems that form in
marginally unstable configurations at radii where interactions with a
planetesimal disk can occur.  At the outset we must recognize that neither
the typical number of massive planets that form in a young planetary
system, nor the typical mass or radial extent of planetesimal disks, is
well constrained by observations. To keep the calculations manageable we
assume here that the properties of the planetesimal disks are ``universal"
(and comparable in terms of mass to the values inferred for the Solar
System), and study how different planetary systems evolve under the joint
action of planet-planet and planetesimal scattering. We fix the number of 
massive planets at $N=3$, but
consider a wide range of planetary mass distributions, including some
modeled after the observed exoplanet mass function and others that
approximate simplified models of the outer Solar System. We follow
the evolution of these systems using an extremely large ensemble of 
N-body simulations, which allows us to reduce the statistical error 
on predicted quantities below that which is currently possible observationally. 

The first results from this project were presented in Raymond \etal (2009b).
In that paper we analyzed a subset of the full set of runs, with the focus
being on the predicted eccentricity distribution and abundance of mean
motion resonances. In this paper we expand our analysis of these topics,
and also extend our study to cover other aspects of planet-planet-disk
interactions. The organization is as follows. In Section 2 we present the
details of the numerical simulations. In Section 3 we study the subset of
simulations that were dynamically unstable: the time to the onset of
instability, typical outcomes, timescales for planet-planetesimal disk
interactions, eccentricity and inclination distributions, and planetary
system packing.  In Section 4 we study mean motion resonances in both
stable and unstable simulations. In Section 5 we discuss how the
eccentricity distribution of extrasolar planets can be understood, both at
small orbital radii (as currently observed) and further out. Finally, we
discuss our results and present our conclusions in Section 6.

\section{Simulations}
Our interpretation of the observations of the outer Solar System, and of
extrasolar planetary systems, is that the typical outcome of the planet
formation process -- shortly after the dispersal of the gas disk --
resembles that shown in Figure~\ref{fig:schematic}. We envisage that two or
more giant planets form, beyond the snow line, with a separation such that
the system will be dynamically active. Beyond the outermost giant planet
lies a region of the disk that successfully formed planetesimals, but which
was unable to form giant planet cores on a timescale comparable to the gas
disk lifetime. At smaller radii lies the zone of terrestrial planet
formation, where again (but for different reasons) planet assembly proceeds
too slowly to allow significant capture of primordial gas. We do not model the
formation of the terrestrial planets in the current work, though the
dynamics of the outer planets can have important effects on their growth.
We also note that while we have motivated these initial conditions
observationally -- by reference to models of exoplanet eccentricities and
the Kuiper Belt -- formation of a system qualitatively akin to
Figure~\ref{fig:schematic} is predicted by standard accretion models.

\subsection{Initial conditions}

\begin{figure}
\centerline{\plotone{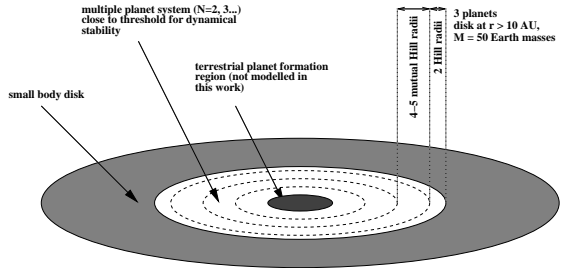}}
\caption{Illustration of the initial conditions of our N-body simulations. 
Conceptually we envisage three radial zones: an inner terrestrial planet 
formation zone, an intermediate region in which gas and ice giants form, 
and an outer disk composed of small bodies that fail to form large planets 
prior to the dispersal of the gas disk. Our specific realization of this 
model assumes that three planets form with an orbital spacing that is close 
to the threshold for dynamical instability in the absence of an exterior disk. 
We additionally assume that the properties of the small body disk are fixed 
across systems, and do not model the terrestrial planet region.}
\label{fig:schematic}
\end{figure}

We adopt a simple and well-defined version of the conceptual model shown in 
Figure~\ref{fig:schematic} as the initial conditions for our scattering 
simulations. We assume that three planets form randomly separated by
4-5 mutual Hill radii $R_{H,m}$,
\begin{equation}
R_{H,m} = \frac{1}{2} \,
(a_1 + a_2) \,
\left(\frac{M_1 +M_2}{3\, M_\star}\right)^{1/3}.
\end{equation}
Here $a$ is the orbital semimajor axis, $M$ is the planetary mass,
$M_\star$ is the stellar mass (fixed at 1 $M_\odot$), and subscripts 1 and
2 refer to the inner and outer planet, respectively.  The spacing of 4-5
$R_{H,m}$, which is common across all but one of our sets of runs, 
was chosen to yield systems that are unstable on a $10^5-10^6$
year timescale (Chambers \etal 1996; Marzari \& Weidenschilling
2002; Zhou \etal 2007; Chatterjee \etal 2008). The exception is our ensemble of runs with 
three $3 \ M_{\rm J}$ planets, for which the number of
unstable cases was so small that we adjusted the spacing to be 3.5-4
$R_{H,m}$ and re-ran the simulations (we still use the set of more
widely-spaced, mostly stable simulations in our analysis of mean motion
resonances in Section~4).
 
The outermost planet of the three planets was placed two (linear) Hill
radii,
\begin{equation} 
R_H = a \left( \frac{M}{3M_*} \right)^{1/3},
\end{equation}
interior to 10 AU. We then chose a single random variable, uniformly
distributed in the range between 4 and 5, to define the interplanetary
spacing. The two additional planets were placed on the appropriate orbits,
moving toward the (Solar-mass) star.  Planets were given zero eccentricity
and randomly-chosen mutual inclinations of less than 1 degree.  We
performed ten sets of simulations, varying the planetary mass distribution
in each set.  For our two largest sets (1000 simulations each) we randomly
selected planet masses according to the observed distribution of exoplanet
masses (Butler \etal 2006),
\begin{equation}
 \frac{{\rm d}N}{{\rm d}M} \propto M^{-1.1}.
\end{equation} 
In the {\tt Mixed1} set we restricted the planet mass $M_p$ to be between a
Saturn mass $M_{\rm Sat}$ and three Jupiter masses $M_{\rm Jup}$.  For our {\tt
Mixed2} set, the minimum planet mass was decreased to 10
$\mearth$.\footnote{In paper 1, the {\tt Mixed1} simulations were referred
to as {\tt highmass} and the {\tt Mixed2} as {\tt lowmass}.}

Although, strictly, we have little knowledge of the actual mass function of
extrasolar planets at orbital radii beyond 5~AU, we regard the runs set up
with the mass function observed at smaller radii as the most
realistic. To gain additional insight into the behavior of more idealized
systems, we performed four additional sets of simulations with equal mass
planets (500 simulations each), and four sets that included radial mass
gradients (250 simulations each). These sets are named with the appropriate
planet masses, starting with the closest to the star outward; for example,
in the {\tt 3J-J-S} set the three Jupiter-mass ($3 M_{\rm J}$) planet is closest
to the star, followed by a Jupiter-mass planet and a Saturn-mass planet.
The eight sets are: {\tt 3J-3J-3J}, {\tt 3J-3J-3J}, {\tt S-S-S}, {\tt
N-N-N}, {\tt 3J-J-S}, {\tt S-J-3J}, {\tt J-S-N}, and {\tt N-S-J}, where
$3J$ refers to a planet mass of $3 M_{\rm J}$, $J$ is $ M_{\rm J}$, $S$ is a Saturn
mass $M_{\rm S}$, and $N$ is $30 \mearth$ (the ``$N$" is intended to refer to
Neptune, although the mass is augmented by just under a factor of two from
Neptune's true mass of 17 $\mearth$ to maintain roughly a factor of three
between the different planet masses).

Each of our $\sim$5000 simulations was run twice: once with just the three
planets and once also including an external planetesimal disk.  The
simulations without disks were presented in Raymond \etal (2008a, 2009a)
and are used in this paper only as a comparison sample.  Simulations with
disks included 1000 planetesimal particles distributed between 10 and 20 AU
following a radial surface density profile $\Sigma \propto r^{-1}$, roughly
consistent with sub-mm observations of outer disks around young stars
(Andrews \& Williams 2007). In all cases the total mass
of the disk was $50 \ M_\oplus$. We note that the inner edge of the disk
lies interior to the radius where a test particle in the restricted 3-body
problem would be stable, so the disk is in immediate dynamical contact with
the outer planet.

\subsection{Integration}
Each simulation was integrated for 100~Myr using the hybrid integrator in
the {\em Mercury} simulation package (Chambers 1999) with a 20 day
timestep.  There is the potential for significant numerical error in the integration for
objects with small perihelion distances, which is usually manifested in
terms of a secular increase in energy for close-in particles (Rauch \&
Holman 1999; Levison \& Duncan 2000).  With our timestep of 20 days, the conventional
wisdom that at least 10 steps are required per orbit implies that we would expect 
to be able to resolve the orbit of a body at 0.67 AU. In fact, a more 
detailed test using a test particle being forced into the star by a giant
planet via the Kozai mechanism shows that the integration error remains
less than $10^{-4}$ down to a perihelion distance of about 0.4 AU.  

We gauged the fidelity of the outcome of each of our simulations based on
the simple energy criterion ${\rm d}E/E < E_X$, where $E_X$ is an energy
threshold.  For the simulations without disks we used the value of Barnes
\& Quinn (2004), who showed that $E_X = 10^{-4}$ is adequate to test for
dynamical stability of multi-planet systems.  For the simulations with
disks, the {\em typical} simulation error was actually ${\rm d}E/E \sim
10^{-4}$, even for cases in which the planets were stable and experienced
very little orbital evolution.  By sifting through a large number of
examples and looking at the energy conservation of individual bodies, it
appeared that an error threshold of $E_X \approx 10^{-3}$ was adequate to
yield reliable results. We therefore set a threshold $E_X = 5 \times
10^{-4}$.  Runs that failed to meet our energy criterion were either rerun
or rejected. For the {\tt Mixed1} and {\tt Mixed 2} sets all runs with
${\rm d}E/E > E_X$ were re-run with smaller timesteps: 5 days for
simulations with no planetesimal disks and 10 days for simulations with
disks. Tests showed that limits of 5 [10] days sufficed to accurately
resolve orbits down to perihelion distances
of 0.15 [0.25] AU.  In each set some of the re-run simulations (typically
15-35) still did not meet our energy criterion and were discarded from the
sample. For the remaining sets of simulations our finite computational
resources did not allow us to re-run those simulations with disks that
failed the energy test. In these cases we discarded the examples with poor
energy conservation based on the initial 20~day timestep runs. We do not
believe that this introduced any substantial bias in our sample because certain
expected outcomes occurred reliably; for example, the eccentricity
distributions for the {\tt J-J-J} and {\tt 3J-3J-3J} simulations are
virtually identical with and without disks.

\subsection{Discussion of the initial conditions}

\begin{figure}
\epsscale{1.25}
\centerline{\plotone{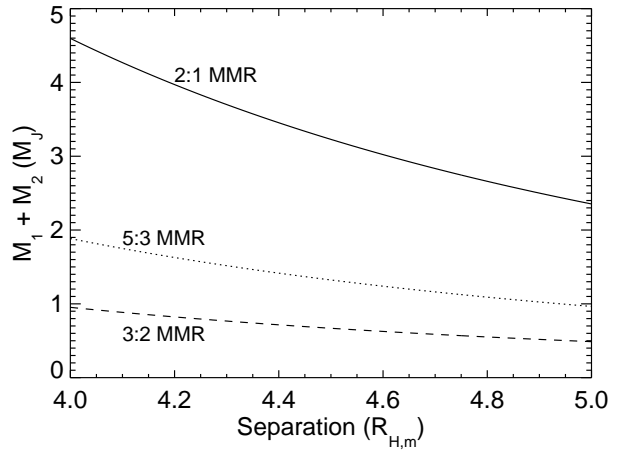}}
\caption{Location of the 2:1, 3:2, and 5:3 mean motion resonances (MMRs)
for two planets in the parameter space of the sum of the two planets'
masses and their separation in mutual Hill radii (see Eq~1). Across a 
range of masses, the initial separations we consider often imply proximity to one 
of these resonances.}
\epsscale{1.0}
\label{fig:mmr-delta}
\end{figure}

The details of our planetary initial conditions are motivated primarily by 
considerations of simplicity, and by the desire to pick planetary 
separations that are not grossly unstable. Our spacing of 
4-5~$R_{H,m}$ would, however, be broadly consistent with models 
in which planets became temporarily trapped in mean motion resonances
(MMRs) during the gaseous disk phase (Snellgrove \etal 2001; Lee
\& Peale 2002; Thommes \etal 2008b). Figure~\ref{fig:mmr-delta} shows the location of the
2:1, 3:2 and 5:3 mean motion resonances (MMRs) as a function of the summed
mass of the planets involved. These are resonances that might be commonly
populated during the gas disk phase, with the 5:3 MMR being generally
unstable once the gas dissipates (Kley \etal 2004; Pierens \& Nelson 2008).  For a range of planet masses between 0.5~$M_{\rm J}$ and
5~$M_{\rm J}$, at least one of these resonances lies in the range probed by our
simulations.  We observe, of course, that a model based specifically on the
idea of resonance trapping followed by release would differ in detail from
ours, since in this case the planetary spacing in units of mutual Hill
radii would correlate with the planetary masses.

Our choices of disk mass and radial extent are also somewhat arbitrary.  A
disk mass of 50~$M_\oplus$ is similar to that used in successful outer
Solar System models, such as the Nice model (Gomes \etal 2005), and it is
also comparable to the summed core masses that one would expect for three
giant planets. It is thus roughly consistent with the idea that
planetesimals in a disk with a continuous surface density distribution
accreted to form cores inside some characteristic radius, while failing to
do so further out. Our inner disk radius of 10~AU, on the other hand, is no
more than a guess. It could typically be larger, in which case {\em
all} of the disk-driven effects discussed in this paper would only be
manifest for planets further out.

\section{Statistical distribution of outcomes}
\begin{deluxetable*}{c|cccc}
\scriptsize
\tablewidth{0pt}
\tablecaption{Scattering Simulations}
\renewcommand{\arraystretch}{.6}
\tablehead{
\colhead{Set} & 
\colhead{N (no disks)} &
\colhead{unstable--frac} &
\colhead{N (disks)} &
\colhead{unstable--frac}}
\startdata
  {\tt Mixed1} &  965 &  569  -- 0.590  &  979  &  442 -- 0.451  \\
  {\tt Mixed2} &  982 &  744  -- 0.758  &  986  &  521 -- 0.528  \\
{\tt 3J-3J-3J}\tablenotemark{1} &  368 &  241  -- 0.655  &  152  &    55 -- 0.362  \\
   {\tt J-J-J} &  452 &  232  -- 0.513  &  380  &   96 -- 0.253  \\
   {\tt S-S-S} &  390 &  362  -- 0.928  &  324  &  196 -- 0.605  \\
   {\tt N-N-N} &  357 &  355  -- 0.994  &  212  &  142 -- 0.670  \\
  {\tt 3J-J-S} &  250 &  150  -- 0.600  &  216  &   55 -- 0.255  \\
  {\tt S-J-3J} &  245 &  219  -- 0.894  &  229  &  147 -- 0.642  \\
   {\tt J-S-N} &  250 &  206  -- 0.824  &  247  &   43 -- 0.174  \\
   {\tt N-S-J} &  245 &  221  -- 0.902  &  177  &  148 -- 0.836  \\
\enddata
\tablenotetext{1}{Recall that we ran an extra set of {\tt 3J-3J-3J}
  simulations (with disks) in which the planets were more closely spaced than our initial set
  of $\sim 500$ simulations (3.5-4 mutual Hill radii as opposed to 4-5).
  The more closely spaced simulations are listed here and used in the
  analysis of Section 3.  The more widely-spaced case contained 498
  simulations of which only two (0.4\%) were unstable -- those simulations
  are used in our analysis of mean motion resonances in Section 4. }
\end{deluxetable*}

Our initial conditions start with planets on orbits that are widely enough
separated that few systems show substantial dynamical evolution on very
short timescales ($10^3$~yr). On longer timescales the dynamical effect of
the disk can be important, and either stabilize or destabilize the system
against close encounters. To analyze our results, we separately consider
the subsets of our runs that were ``stable" versus those that were
``unstable". In this context, we define a stable system as one in which
{\em no close encounters} between planets occurred over 100~Myr.  A close
encounter occurs when any two planets enter their mutual Hill sphere. An
unstable system is thus one in which any two planets underwent at least one
close encounter. Table~1 lists the fraction of each set of simulations that
was stable with and without disks.

The above definition of stability, while precise and easily measured from
the simulations, fails to capture the full range of behavior seen in runs
with planetesimal disks. We therefore introduce a different classification
in \S3.3, where we consider how the characteristic timescale for evolution
of the joint planet-disk system correlates with the presence of close
encounters and architectural re-arrangement.

\subsection{Instability Timescales}

\begin{figure*}[t]
\centerline{\plotone{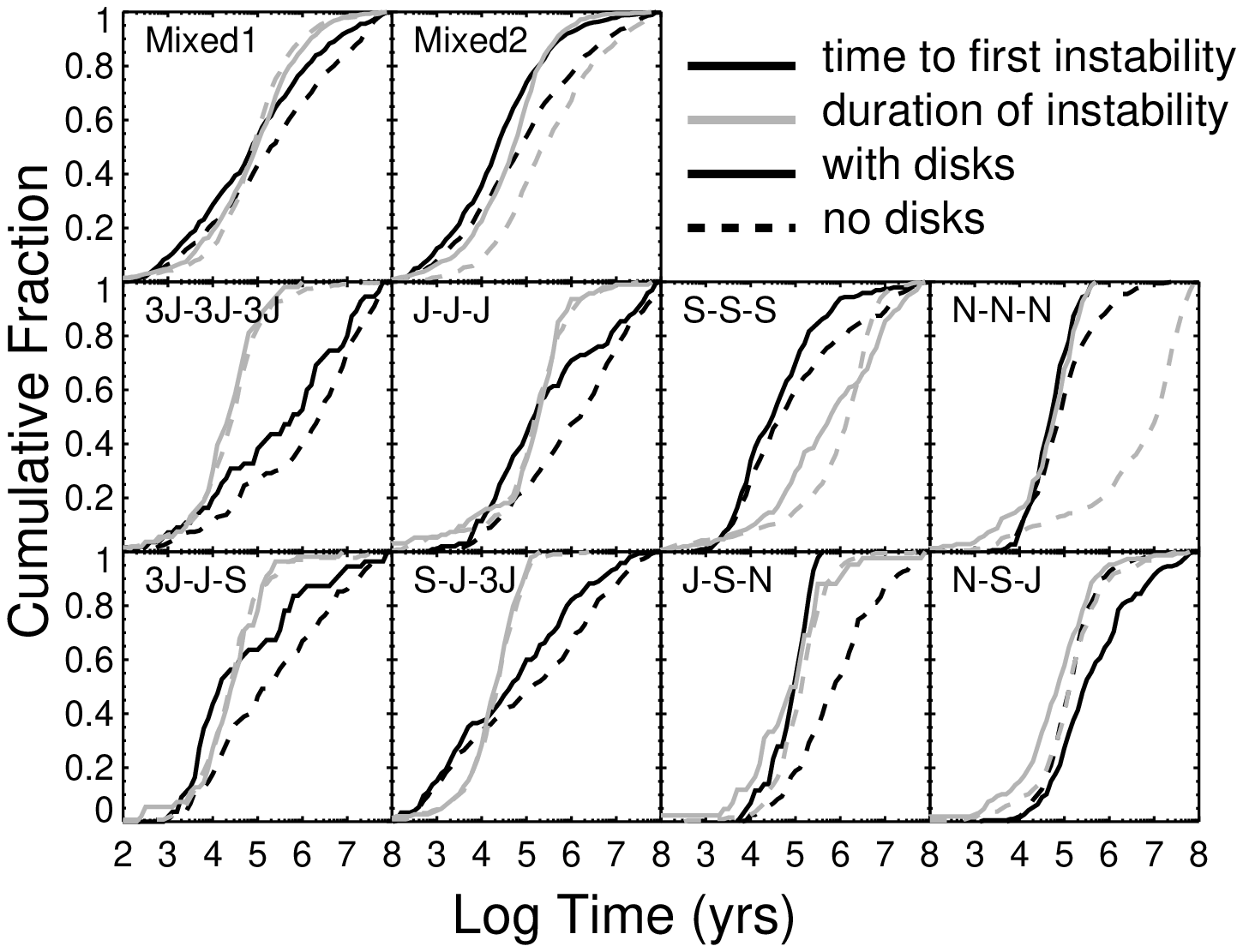}}
\caption{Cumulative distributions of instability timescales for each set of
simulations.  Solid lines represent cases with disks, and dashed lines
without disks.  Note that the total planet mass increases to the left.}
\label{fig:tinst}
\end{figure*}

Given the spacing of 4-5 mutual Hill radii, we expect that the typical
instability timescale in our simulations should be $10^5 - 10^6$ years
(Mazari \& Weidenschilling 2002; Chatterjee \etal 2008).  Indeed, the
median time for the first close encounter between planets in the cases
without disks ranged from 58,000 years ({\tt S-S-S}) to 3.1 Myr ({\tt
3J-3J-3J}).  Instabilities typically set in earlier for systems containing
less massive planets (see Chambers \etal 1996), and when the planet mass
increases outwards (i.e., {\tt N-S-J} simulations went unstable more
quickly than {\tt J-S-N} simulations).

Dynamically unstable systems tend to stabilize by destroying one or more
planets, usually via collision with another planet or hyperbolic ejection.
Systems with equal-mass planets tend to undergo far more close encounters
over a much longer timespan than systems with significant mass differences
between planets. The planets that survive in systems with
equal-mass planets have much larger eccentricities than those in
unequal-mass systems (Ford \etal 2003; Raymond \etal 2008a). 
The duration of instability is also mass-dependent: systems with low but 
equal-mass planets remain unstable for longer periods than systems in which 
the planet masses are scaled up. 

Figure~\ref{fig:tinst} shows the cumulative distribution of (1) the timescale 
from the start of each simulation to the first close encounter between planets 
(shown in black), and (2) the duration of the system instability, i.e. the 
time from the first to the last close encounter. We plot all ten sets of runs, 
both with and without planetesimal disks. For this unstable subset of runs, 
the 100~Myr duration of the simulations is sufficient to observe the onset and 
full duration of the instability in almost all cases. 

With the sole exception of the {\tt N-S-J} runs, the onset of instabilities
in systems with disks occurs sooner than in those without disks
(Fig.~\ref{fig:tinst}). One important effect of the presence of the
planetesimal disks is clearly to stabilize systems that might otherwise be
unstable on long timescales.  This effect is most pronounced in systems
with low-mass outer planets (e.g., the {\tt J-S-N} cases). The reason for
this is likely related to the mass dependence of the migration rate of a
planet through a planetesimal disk, which is a strongly decreasing function
of planet mass once the mass exceeds the mass of planetesimals within a few
Hill radii (Ida \etal 2000; Kirsh \etal 2009). This migration can cause two
planets (usually the outer two) to cross a low-order mutual mean motion
resonance (see Fig~\ref{fig:mmr-delta}). MMR crossings lead to an increase
in planetary eccentricities (e.g., Chiang \etal 2002; Tsiganis \etal 2005),
and trigger an instability that leads to close encounters.

Disks also reduce the duration of the unstable phase, but only in systems
containing lower-mass outer planets. The most dramatic example are the {\tt
N-N-N} simulations, for which the typical period of close encounters was
reduced by a factor of more than 100 (Fig.~\ref{fig:tinst}).  This occurs
because the orbits of high-eccentricity planets are circularized via
dynamical friction with the planetesimal disk (e.g., Thommes \etal 1999),
and this process is faster and more efficient for lower-mass planets.  As
scattered planets are re-circularized by the disk, their periastron
distances are increased, they are removed from the region of scattering and
future encounters between planets do not occur.  In contrast, the planetary
dynamics in systems with a massive planet adjacent to the planetesimal disk
are effectively shielded from the effects of the planetesimal disk such
that the duration of instabilities is the same as for simulations with no
disks.

\subsection{Characteristic Evolution}
In planetary systems with no dissipation, once an instability occurs the
endpoint is almost inevitably a violent event: a planet-planet collision, a
planet-star collision, or a hyperbolic ejection from the
system\footnote{Ford \etal (2001) found that ``quasi-stable''
configurations could arise in which no planet was destroyed but the orbits
were chaotic.  We also found several such systems, but each one ended up
being unstable if integrated for a long enough interval.  It is conceivable
that a small fraction of systems may end up on such orbits and be observed
before they become unstable.}.  However, in systems with dissipation from
gaseous or planetesimal disks, the situation changes and the number of
potential outcomes increases.  The key new processes that arise from the
presence of a planetesimal disk occur primarily for planets whose masses
are similar to the planetesimal disk mass.  Given our disk mass of 50
$\mearth$, it is the simulations that include at least one roughly
Saturn-mass or smaller planet that are affected.  Interactions between the
planetesimal disk and lower-mass planets can lead to angular momentum
exchange and radial migration (Fernandez \& Ip 1984). In addition,
dynamical friction from the planetesimal disk can re-circularize the orbit
of a scattered planet.  Both of these processes feature prominently in
recent models of the dynamical evolution of the Solar System's giant
planets (Malhotra 1995; Thommes \etal 1999; Tsiganis \etal 2005; Ford \& Chiang 2007).

\begin{figure*}
\centerline{\plotone{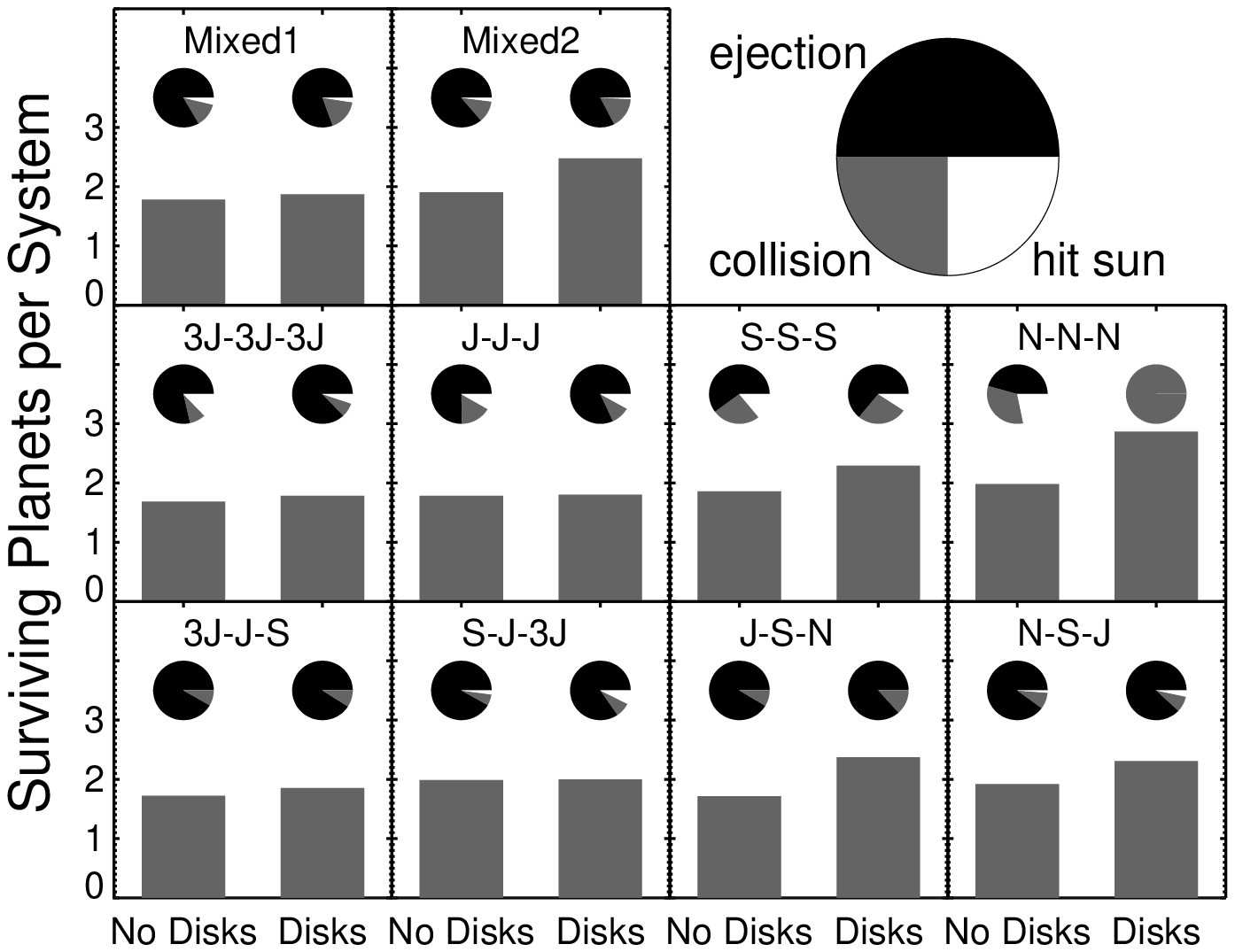}}
\caption{A summary of the outcomes of our simulations.  The bar graph in
each panel shows the mean number of surviving planets per system for all
cases that were unstable, with (right) and without (left) disks.  The pie
chart in each panel shows what happened to the destroyed planets in each
case: ejection (black), planet-planet collision (gray), or collision with
the star (white).}
\label{fig:outcomes}
\end{figure*} 

Figure~\ref{fig:outcomes} shows the range of outcomes of the unstable
simulations for each giant planet configuration.  The number of planets
surviving per system increases for simulations with disks, but only for
cases that contain a low-mass planet.  The most dramatic difference is seen
in the {\tt N-N-N} simulations, for which the number of planets per
unstable system increased from 1.98 to 2.87, meaning that only one out of
every 7.5 unstable simulations with disks destroyed a planet.  The pie
charts in Fig.~\ref{fig:outcomes} show what happened to the destroyed
planets in each case.  For almost all cases, the dominant destruction
mechanism was ejection from the system, which accounted for at least 3/4 of
the destroyed planets in all configurations except {\tt S-S-S} and {\tt
N-N-N}.  In the {\tt N-N-N} simulations without disks, ejection and
collisions played a comparable role, which is not surprising given that the
ratio of the escape speed from the planet to the escape speed from the
planetary system (sometimes called the ``Safronov number'', $\Theta$), is
the smallest for any system, only about 2 (Ford \etal 2001; Goldreich \etal 2004).  In the
{\tt N-N-N} simulations with disks, gravitational kicks during close
encounters were not strong enough to eject a single planet in any
simulation.  The reason for this is twofold.  First, dynamical friction
efficiently circularizes the orbits of low-mass planets such that the
typical encounter speed is low.  Second, dynamical friction is also
effective at effectively capturing scattered planets in the disk since the
total disk mass actually exceeds the planet mass (30 $\mearth$) in this
case.

\subsection{Timescales for Planet-Planetesimal Disk Interactions}

We have seen that the planetesimal disk can have a strong influence on the
planets' dynamical evolution.  The planets have a back-reaction on the disk
itself, pumping up planetesimal eccentricities via dynamical friction and
removing planetesimals from the system via dynamical ejection or collision.
The number and strength of planet-planetesimal interactions depends most
strongly on the planetary orbits.  If the planets' orbits remain circular
then the planets can interact with only a fraction of the disk and
encounter velocities depend mainly on the planetesimal eccentricities.  If,
however, planetary orbits are eccentric then the planets can interact with
a larger fraction of the disk and encounter speeds will be larger,
resulting in more frequent ejection of planetesimals.  Thus, the
perturbations felt by the planetesimal disk will correlate with the
planetary orbits and therefore the dynamical stability of the planets. 

\begin{figure}
\epsscale{1.2}
\centerline{\plotone{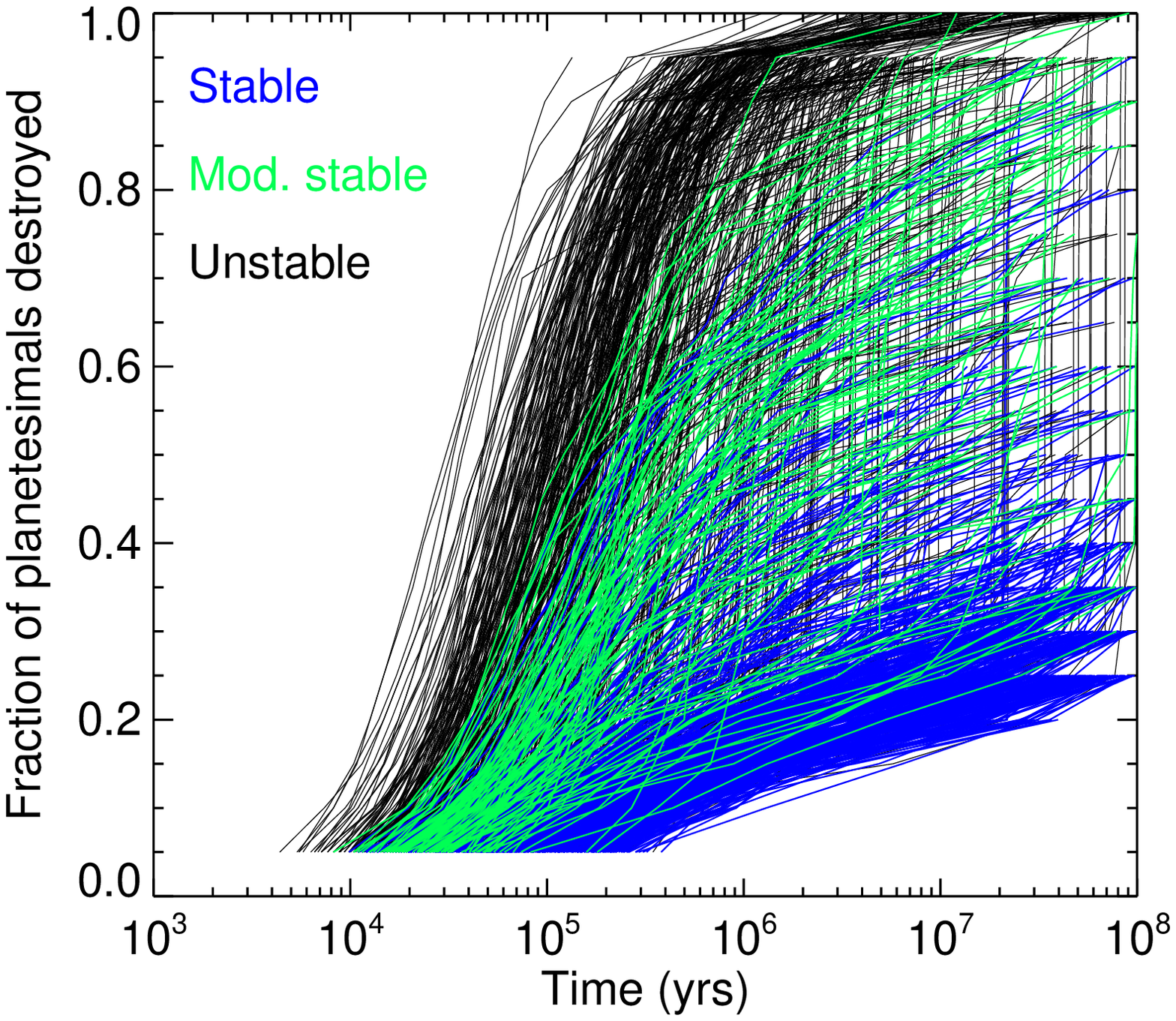}}
\caption{Timescale for the destruction (via hyperbolic ejection, collision
with a planet or collision with the central star) of the planetesimal disk
for the {\tt Mixed1} simulations.  Black lines correspond to unstable
simulations which had at least one close encounter between planets.  Green
lines refer to ``moderately stable'' simulations which experienced no
planetary close encounters but the final planetary system had a
mass-weighted eccentricity larger than 0.025.  Blue lines correspond to
``stable'' simulations, which experienced no close encounters between
planets and remained on low-eccentricity orbits.  Note that we only plot
every 5\% of planetesimals destroyed.}
\label{fig:plandest1}
\epsscale{1}
\end{figure}

Figure~\ref{fig:plandest1} shows the timescale for the destruction of a
given fraction of the planetesimal disk for all of the {\tt Mixed1}
simulations, where each curve corresponds to a single simulation.  Here we
have subdivided the ``stable" simulations, which never experience close
encounters, into moderately stable and stable subsets. We define a
moderately stable system as one which never experienced a close encounter
between planets but for which the final mass-weighted planetary
eccentricity is larger than 0.025. These systems have thus experienced
significant planet-planet perturbations despite the absence of close
planetary encounters.

The unstable {\tt Mixed1} systems destroy the majority of their
planetesimal disks on a $10^5-10^6$ year timescale
(Fig.~\ref{fig:plandest1}), although late-onset instabilities can be seen
as nearly vertical lines at later times.  There is often a delay for the
destruction of the last 10\% of planetesimals because in most cases these
have large inclinations such that the close encounters with the planets
needed to reach zero energy are less frequent.  In fact, many unstable
cases do not destroy the entire planetesimal disk.  Moderately stable
systems can have a range in the timescale and amount of planetesimal
destruction depending on the planet masses and evolution.  For relatively
massive planets, the sudden eccentricity increase that accompanies a
resonance crossing leads to a corresponding jump in the eccentricities of
many or most disk particles, leading to orbit crossings, close encounters
and ejection.  For less massive planets secular perturbations are weaker so
less of the disk is destabilized during such events.  One configuration
that is quite efficient at clearing out planetesimals is a lower-mass outer
planet and a high-mass middle planet.  The outer planet's low mass allows
it to migrate outward somewhat due to planetesimal scattering.  The
scattered planetesimals, in turn, are quickly ejected by encounters with
the high-mass middle planet.

The stable {\tt Mixed1} systems, which are relatively high-mass, destroy
only a relatively small portion of the planetesimal disk.  The outer planet
interacts directly with planetesimals within the stability boundary and
also may excite planetesimals that happen to be in resonances onto
planet-crossing orbits.  These directly-interacting planetesimals can be
ejected by the outer planet or can be ``passed'' inward to interact with
the inner planets, which typically eject the planetesimals.  Thus, the
inner planetesimal disk is quickly cleared out and interactions between the
planets and the planetesimal disk become infrequent, with 50-75\% of the
disk remaining.

The evolution of the planetesimal disks in our simulations offers a rich
data set to understand the connection between planetary dynamics and disk
structure.  In a future paper, we will correlate the planetary and
planetesimal components of each system and calculate the infrared
detectability of these disks for each case.  For the remainder of this
paper, we restrict ourselves to the planetary dynamics.

\subsection{Eccentricity Distributions}

\begin{figure*}
\centerline{\plotone{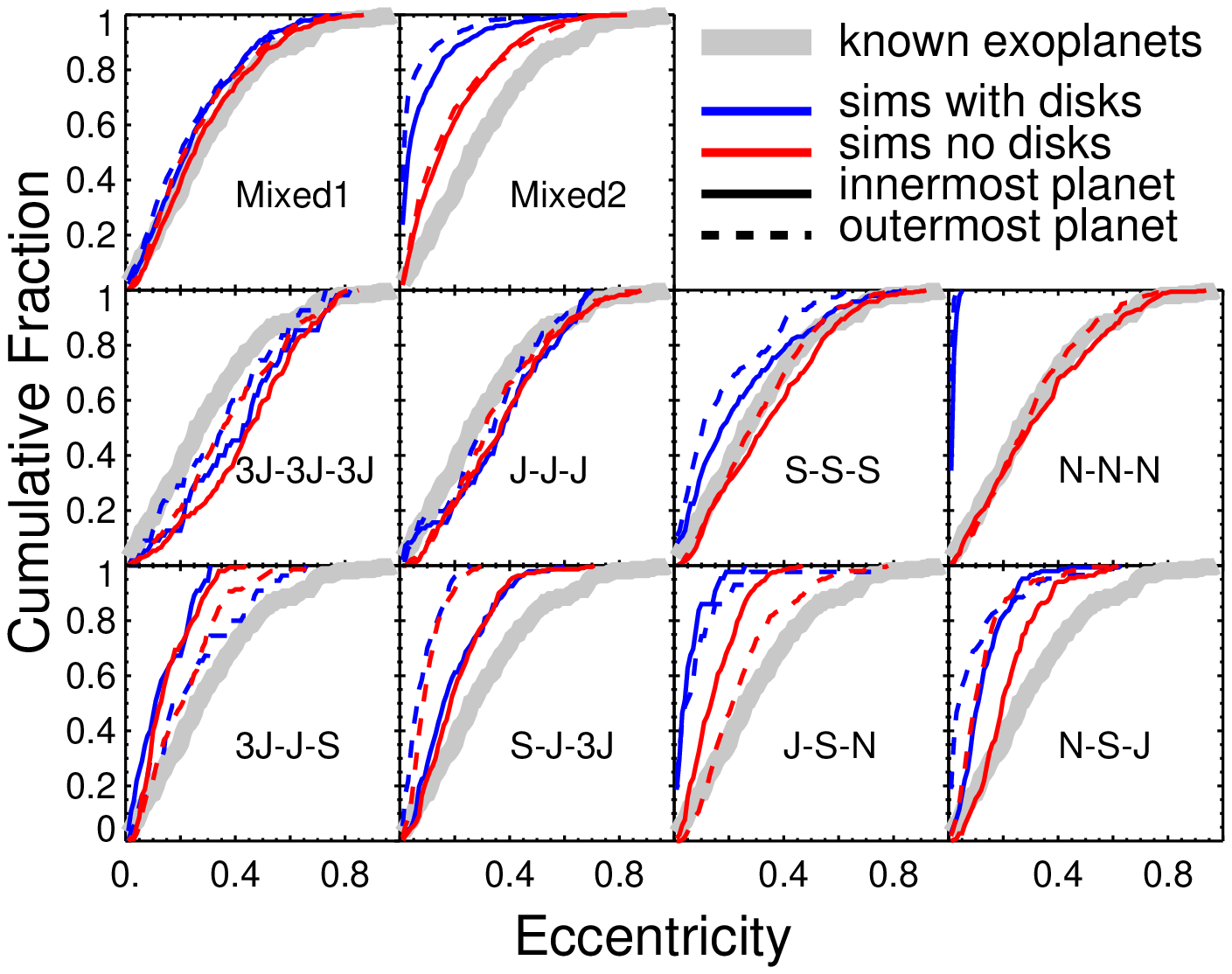}}
\caption{Cumulative eccentricity distributions for the unstable simulations
from each of our ten simulation cases, with (blue) and without (red)
planetesimal disks, compared with the observed extra-solar systems exterior
to 0.1 AU (thick grey line).  The innermost planet from each simulation is
shown with the solid lines, while the outermost planet is shown with the dashed lines.}
\label{fig:ecum_all}
\end{figure*}

Figure~\ref{fig:ecum_all} shows the cumulative eccentricity distributions
for the innermost and outermost planets in all of our unstable simulations,
with and without disks. We also plot the observed exoplanet eccentricity
distribution, excluding planets inside 0.1 AU which are likely to have had
their orbits altered by tides (Jackson \etal 2008). We note that, since our
simulations start with planets at fairly large radii, we do not populate
the full radial range across which the observed distribution is
determined. Currently, however, the evidence for any radial dependence to
the eccentricity distribution is marginal (Ford \& Rasio 2008), so the
comparison between the observed and simulated distributions is justifiable.

Although the result is now well-established (Chatterjee \etal 2008; Juri\'c
\& Tremaine 2008), it is still startling to see from
Fig.~\ref{fig:ecum_all} how easily planet-planet scattering can reproduce
the observed eccentricity distribution.  With just the simple assumption
that planets have the observed mass distribution and form on marginally
unstable orbits (i.e., the {\tt Mixed1} simulations), we can produce a
statistical match to the observed distribution.  The planet-planet
scattering model appears to be a very robust and simple way to explain the
observed systems.

As expected, the difference between planetary eccentricities for
simulations with and without disks is strongly mass dependent and clearly
seen in a comparison between the {\tt Mixed1} and {\tt Mixed2} simulations.
The effect of the planetesimal disk is clear for $M_p \lesssim M_{\rm S}$, and
for the lowest-mass case, {\tt N-N-N}, orbital re-circularization from
planet-planetesimal disk interactions is so strong that there are no
high-eccentricity planets at all -- the most eccentric planet in all the
{\tt N-N-N} simulations with disks is just 0.06.  As seen in previous work,
scattering among equal-mass planets produces larger eccentricities than
scattering among planets with different masses (Ford \etal 2003; Raymond
\etal 2008a), and this is clearly seen in Fig.~\ref{fig:ecum_all}.

Variations between the eccentricities of inner and outer planets are caused
by two different effects: scattering from other planets and eccentricity
damping from the planetesimal disk.  For the {\tt Mixed1} simulations there
is very little difference between the inner and outer eccentricity
distributions because there is no planetary mass gradient and the planet
masses are too large to be significantly affected by the planetesimal disk.  
For the {\tt Mixed2} simulations with
disks, outer planets have lower eccentricities. This offset is not seen for the
simulations without disks.  Thus, the lower-eccentricity outer planets are due
to damping from the planetesimal disk.  For simulations with equal-mass
planets, there is a modest inner vs. outer eccentricity difference because
of disk damping but only for the {\tt S-S-S} cases, as the Jupiter-mass
planets ({\tt J-J-J}) barely feel the disk and 30 $\mearth$ planets ({\tt
N-N-N}) are all quickly circularized.  For systems with radial gradients in
planet mass there is an inner vs. outer eccentricity difference caused by
the simple mass-dependence of the recoil velocity during a planetary
encounter.  During instabilities involving different-mass planets, this
leads to more massive planets having smaller eccentricities than less
massive planets.  For systems with negative mass gradients ({\tt 3J-J-S}
and {\tt J-S-N}), this causes the outer planets to have higher
eccentricities than the inner planets, and the effect is reversed for
systems with positive mass gradients ({\tt S-J-3J} and {\tt N-S-J}).  For
the lower-mass cases with mass gradients ({\tt J-S-N} and {\tt N-S-J}), the
effects of preferential eccentricity damping of outer planets by the
planetesimal disk are also clearly seen.

\begin{figure}
\centerline{\plotone{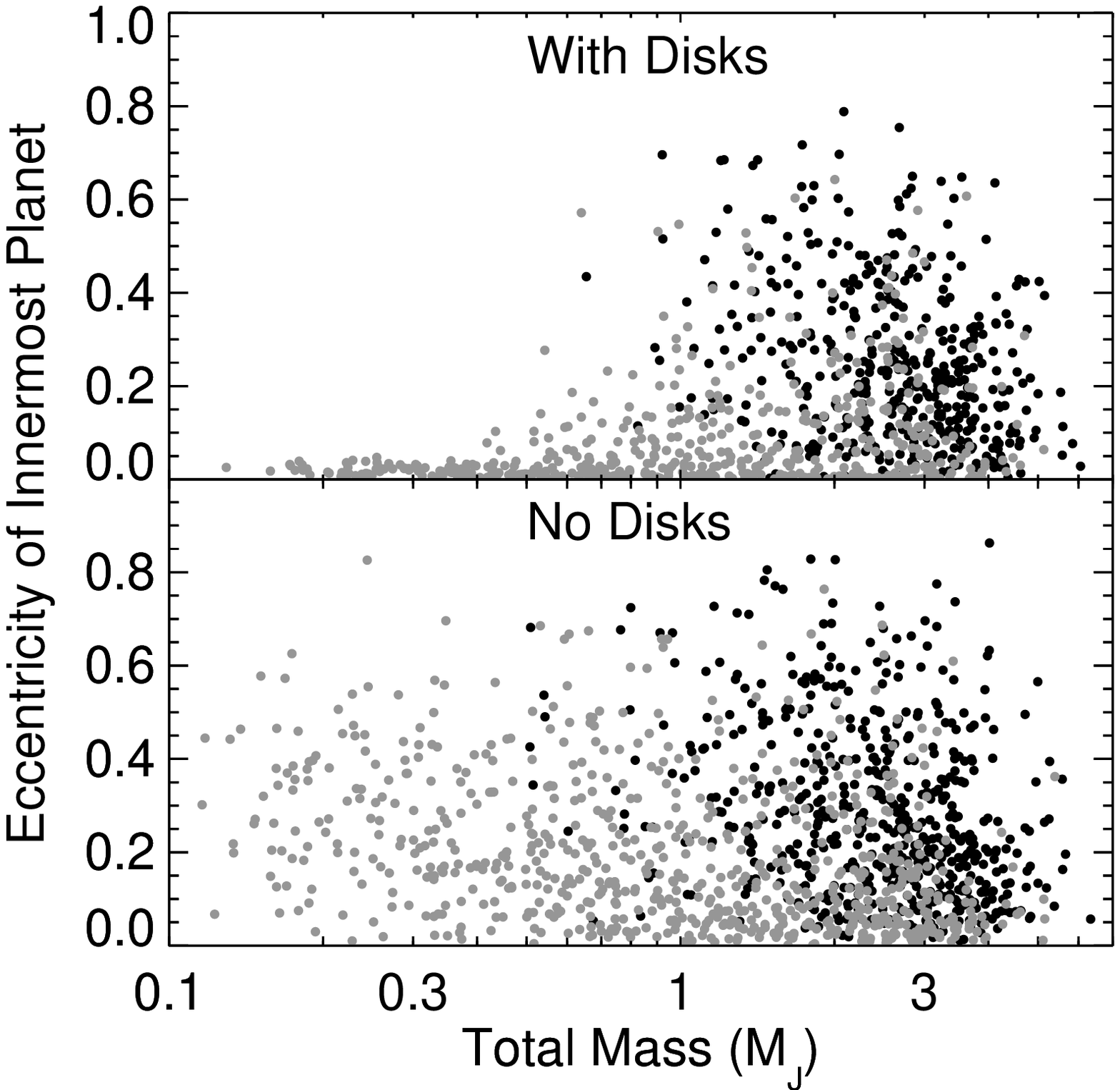}}
\caption{Final eccentricity of the innermost planet vs. the
  total mass in surviving planets for the unstable {\tt Mixed1} (black) and
  {\tt Mixed2} (grey) simulations, with (top panel) and without (bottom
  panel) planetesimal disks.  Disks result in a sharp transition to a
  low-eccentricity regime for system masses below about one Jupiter mass. 
  For higher system masses, the observed exoplanet eccentricity distribution 
  is recovered irrespective of the presence or absence of disks.}
\label{fig:e-mtot}
\end{figure}

\begin{figure}
\centerline{\plotone{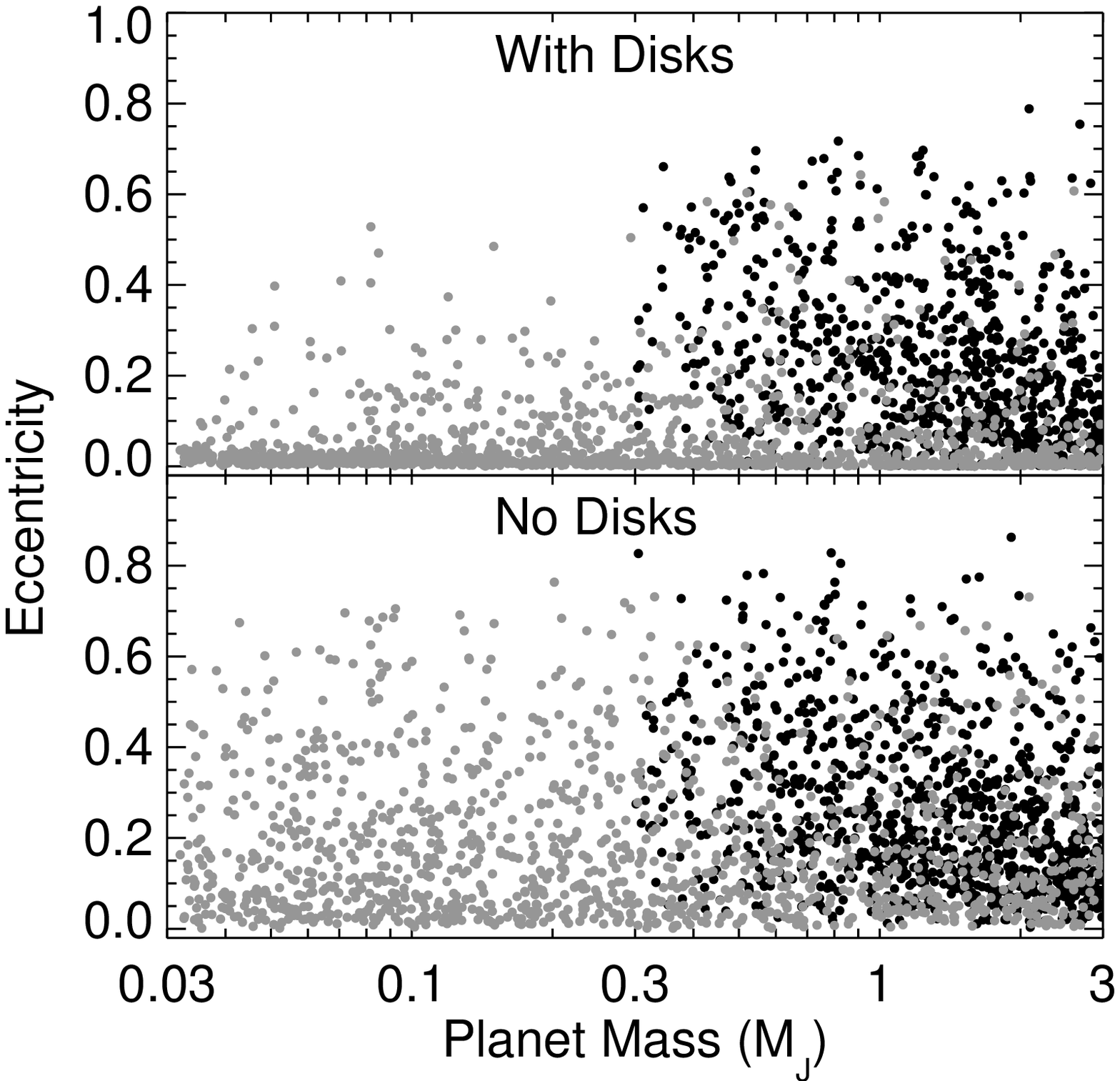}}
\caption{Planetary eccentricity vs. mass for all surviving planets in the
  unstable {\tt Mixed1} (black dots) and {\tt Mixed2} (grey dots), with
(top panel) and without (bottom panel) planetesimal disks. }
\label{fig:e-m}
\end{figure}

In Figure~\ref{fig:e-mtot} we show the eccentricity of the innermost (and
thus most easily detectable) planet as a function of the final total mass
of the planetary system, using just the {\tt Mixed1} and {\tt Mixed2}
simulations. In the presence of disks there is a strong correlation between
these quantities (Raymond \etal 2009b). For systems with total masses of
less than $\sim 1 \, M_{\rm J}$, circular or near-circular orbits dominate. 
For higher system masses, on the other hand, planetary eccentricities
display a large range, from near zero to $> 0.8$. This correlation is much
weaker if we plot, instead, planetary eccentricity against individual
planet mass for all surviving planets (Figure~\ref{fig:e-m}). As is obvious
from the cumulative distributions, disks do act to circularize low mass
planets more than high mass planets, but there is no sharp transition
between the two regimes. We interpret this difference as being due to the
fact that it is the total mass of the planetary system that determines the
disk lifetime and the ability of the disk to damp eccentricities, rather
than any individual planet mass (see discussion in \S 3.6 below).

\subsection{Detectability of a transition to low $e$ at large orbital radii}
\begin{figure}
\centerline{\plotone{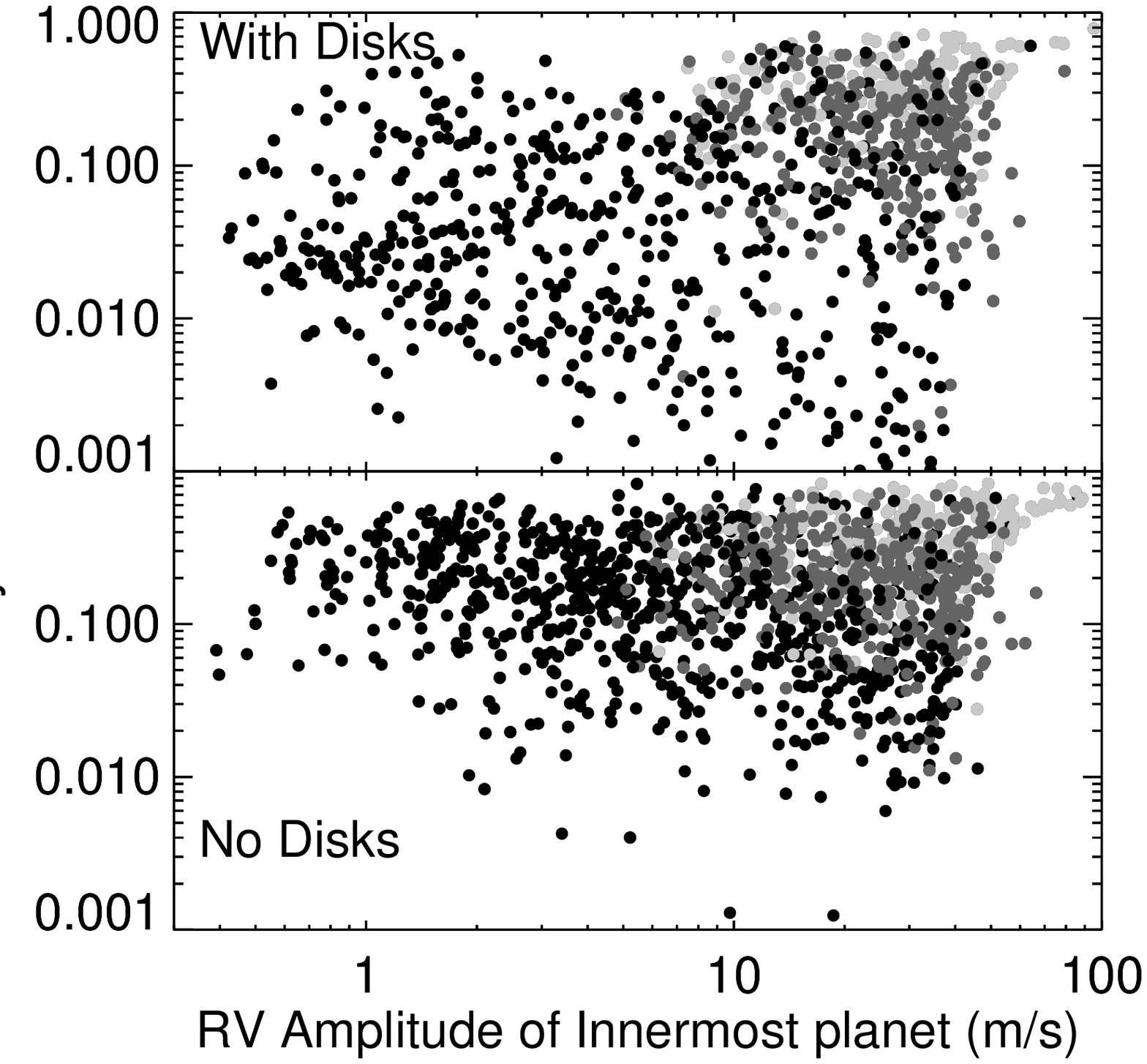}}
\caption{The predicted signature of the transition to low 
eccentricity orbits in radial velocity surveys. The eccentricity of the 
innermost surviving planet is plotted as a function of the stellar 
radial velocity amplitude $K$ for the {\tt Mixed1} and {\tt Mixed2} 
simulations both with and without disks. The light, intermediate, and 
dark grey points correspond to planets with orbital periods $P < 2000 \ {\rm days}$, 
$2000 \ {\rm days} < P < 4000 \ {\rm days}$, and $P > 4000 \ {\rm days}$, respectively. 
Detection of the dynamical influence of planetesimal disks requires either 
long duration (in excess of 10~years) monitoring sensitive to planets with 
$K < 10 \ {\rm m \ s}^{-1}$, or highly accurate determinations of the 
eccentricity of planets on almost circular orbits.}
\label{fig:RV}
\end{figure}

The semi-major axis, beyond which our model predicts a transition to 
low eccentricity orbits for $M < M_{\rm J}$ planets, depends upon the 
characteristic inner radius of the primordial planetesimal disk 
(note that the {\em mass} at which the transition occurs will also vary with 
this quantity). 
This is a free parameter, which we have fixed at 10~AU. If, in 
reality, planetesimal disks do typically extend in to about 10~AU, 
our results suggest that the transition to the low eccentricity 
regime may be observable with feasible extensions of existing 
radial velocity surveys (if the inner extent of planetesimal disks 
is at substantially larger radii astrometry or direct imaging may offer 
better prospects). To quantify this, we have calculated the radial 
velocity amplitude $K$, 
\begin{equation}
 K = \frac{1}{\sqrt{1-e^2}} \left( \frac{M}{M_*} \right) \sqrt{\frac{GM_*}{a}} \sin i,
\end{equation}
for the innermost surviving planet for all of the {\tt Mixed1} and {\tt Mixed2} runs.
The distribution of $e$ as a function of $K$ is shown in Figure~\ref{fig:RV}, assuming 
$M_* = M_\odot$ and (for simplicity) that $\sin i = 1$. Since the detectability 
of a planet via radial velocity measurements depends upon the period as well 
as the amplitude, the points are further coded to indicate relatively short 
period ($P < 2000 \ {\rm days}$), intermediate period ($2000 \ {\rm days} < P < 4000 \ {\rm days}$), 
and long period ($P > 4000 \ {\rm days}$) planets.

A measurement of $e(K)$ for the innermost planet provides no information as 
to the total mass of the planetary system, and as a consequence the transition 
to lower eccentricities at lower $K$ resembles that shown in Figure~\ref{fig:e-m} 
($e$ as a function of individual planet mass) rather than the sharper 
transition seen in Figure~\ref{fig:e-mtot} ($e$ as a function of the final total 
system mass). Nonetheless, we see a clear trend to lower eccentricities 
that sets in for radial velocity amplitudes $K \lesssim 10 \ {\rm m \ s}^{-1}$. 
Although there are high $e$ outliers at all values of $K$, for $K = 1-5 \ {\rm m \ s}^{-1}$ 
the typical eccentricity lies in the $10^{-2} - 0.1$ range, whereas for 
$K > 10 \ {\rm m \ s}^{-1}$ values $e > 0.1$ are obtained. Comparing the results 
with and without disks makes clear that this difference is caused by the 
damping effects of planetesimals.

Detection of planets with radial velocity amplitudes $K < 10 \ {\rm m \ s}^{-1}$, 
although not easy, is already possible (almost 40 such systems are currently known). 
The primary obstacle to observational study of the outer planet region, where the 
dynamical effects of planetesimal disks should become evident, is rather the long 
periods of planets. As is clear from Figure~\ref{fig:RV}, essentially all of the 
low mass planets, whose orbits have been partially circularized by interaction 
with planetesimals, have periods in excess of 4000~dy. High precision radial 
velocity measurements, capable of finding planets with $K \approx 5 \ {\rm m \ s}^{-1}$ 
and periods in excess of 10~years, are therefore required in order to start 
probing the dynamics discussed in this paper. 

Inspection of Figure~\ref{fig:RV} suggests a second observational signature of 
planetesimal disk dynamics: the existence of {\em extremely} low eccentricity 
planets. In the presence of disks, a small but measureable fraction of even 
massive planets (with $K = 10-30 \ {\rm m \ s}^{-1}$) end up with $e < 10^{-2}$, 
whereas such systems are rare in the absence of disks. Extremely precise 
measurements of eccentricity, capable of detecting an excess (above the predictions 
of pure scattering models) of truly circular orbits, would therefore be 
valuable even if such observations were only available for massive planets. 
The interpretation of an excess of circular orbits, however, would likely 
be more ambiguous than a measurement of the full $e(K)$ distribution, since it 
could reflect small amounts of gas damping or an admixture of systems that 
only formed one planet, as well as the dynamical influence of planetesimals.

\subsection{Inclination Distributions}

Planet scattering can lead to large mutual inclinations between the
surviving planetary orbits, as well as misalignment with respect to the initial plane of
the planets (Chatterjee \etal 2008; Juri\'c \& Tremaine 2008).  Mutual
inclinations between the orbital planes of planets in multi-planet systems
can be detected via astrometry (e.g., Bean \& Seifahrt 2009), while 
any misalignment between the orbital plane and the plane perpendicular 
to the stellar spin axis can be detected directly for
transiting planets via the Rossiter-McLaughlin effect (Gaudi \& Winn 2007;
Winn \etal 2005).  Another interesting consideration is that the initial
planetary plane coincides with the plane of the planetesimal disk and
presumably also with the plane of the dust disk that should be produced by
collisional grinding of planetesimals (e.g., Wyatt 2008).  

\begin{figure*}
\centerline{\plotone{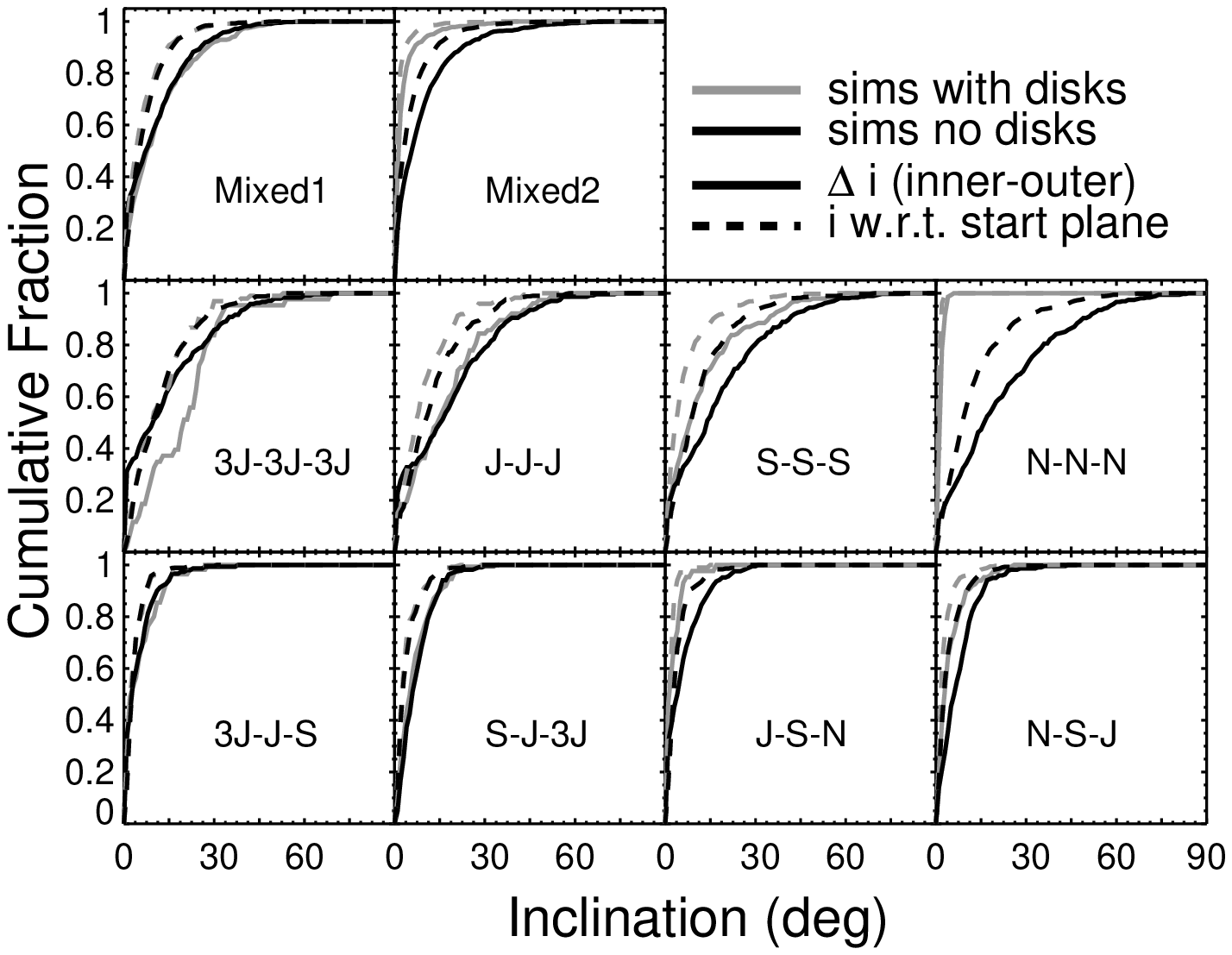}}
\caption{Cumulative inclinations distributions for the unstable simulations
from each of our ten simulation cases, with (blue) and without (red)
planetesimal disks.  The solid lines show the distribution of the mutual
inclination between the orbits of the inner and middle planets for systems
with two surviving planets.  The dashed lines show the distribution of
inclinations with respect to $i=0^\circ$, the initial plane of the planets
and planetesimals.  }
\label{fig:icum}
\end{figure*}

Figure~\ref{fig:icum} (dashed lines) shows the distribution of inclinations
with respect to the initial orbital plane for the unstable systems in all
ten sets of simulations, both with and without planetesimal disks.  As was
the case for eccentricities, larger inclinations are generated in systems
with equal-mass planets than in systems with large mass ratios.
Inclinations in the few to $\sim$15 degree range are common in most sets of
simulations, but only the equal-mass systems are able to generate larger
inclinations.  Unlike the eccentricity distributions for which higher-mass
planets have higher eccentricities, all four equal-mass systems (without
disks) have virtually identical inclination distributions.  

\begin{figure}
\centerline{\plotone{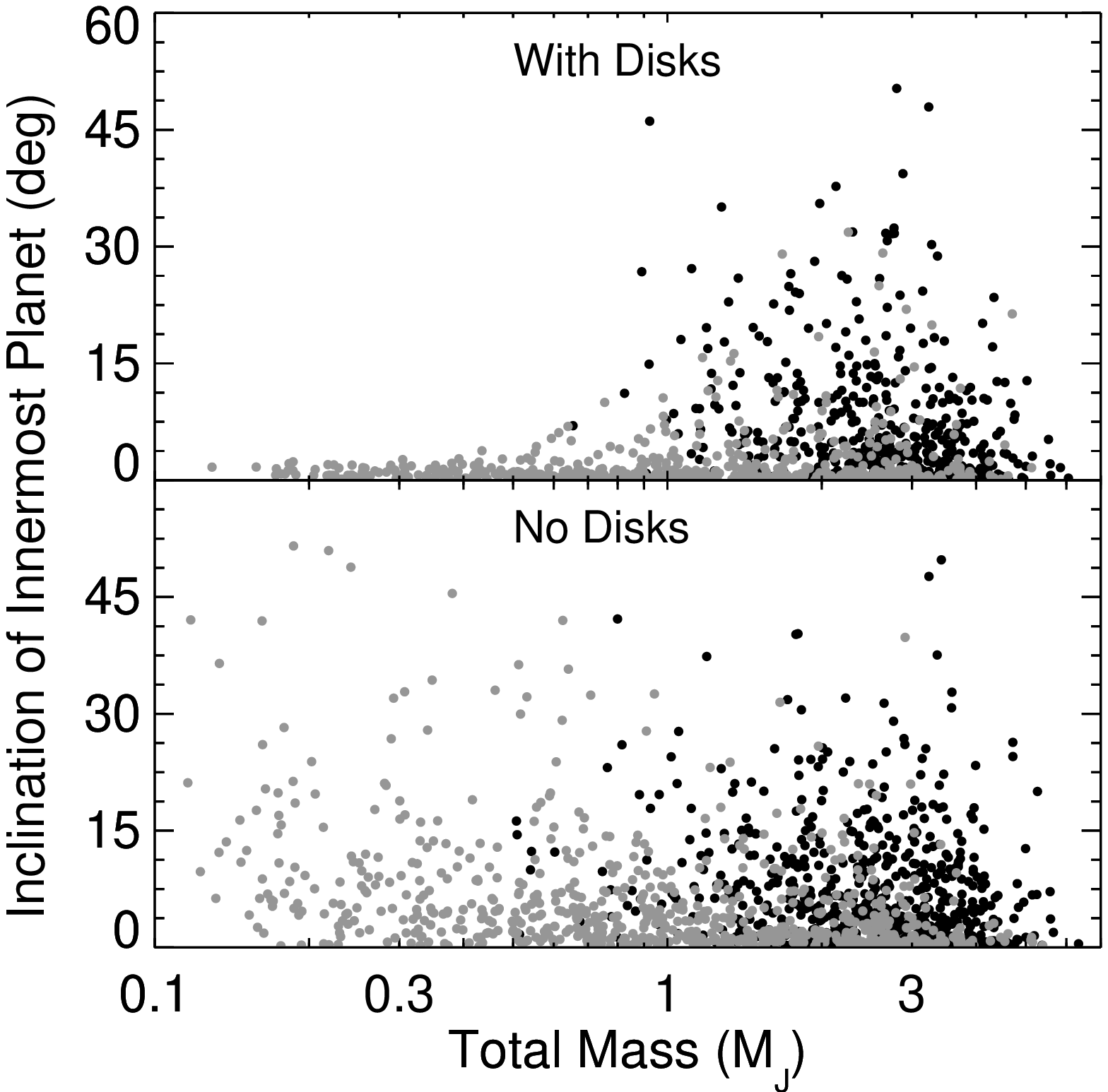}}
\caption{The inclination of the innermost planet with respect to the
initial orbital plane as a function of the total mass in surviving planets
for the {\tt Mixed1} (black dots) and {\tt Mixed2} (grey dots) simulations. }
\label{fig:mtot-iin}
\end{figure}

As expected, the simulations with disks yielded smaller inclinations than
the simulations without disks, and the effect was stronger for lower-mass
planets. Figure~\ref{fig:mtot-iin} shows the inclination of the
innermost planet, which as in the case of eccentricity is controlled by the
total planetary mass rather than the individual planet masses.  The same
sharp break at $M_{tot} \approx 1 M_{\rm J}$ is seen to divide planets with small
inclinations vs. those with a large range in inclination.

The mutual inclination $\Delta i$ between two planetary orbits with
inclinations $i_1$ and $i_2$ is given by:
\begin{equation}
\cos{\Delta i} = \cos{i_1} \cos{i_2} + \sin{i_1} \sin{i_2}
\cos{\left(\Omega_1-\Omega_2\right)}, 
\label{eq:inc}
\end{equation}
\noindent where $\Omega_1$ and $\Omega_2$ refer to the longitudes of
ascending node.  Figure~\ref{fig:icum} shows the distributions of the
mutual inclination $\Delta i$ between the innermost and the adjacent planet
in unstable systems for which two or more planets survived (solid lines).
The $\Delta i$ values are consistently larger than the inclinations with
respect to the initial orbital plane.  During a close planetary encounter,
if one planet is scattered in the $\hat{z}$ direction, then the other
planet must receive a kick in the $-\hat{z}$ direction.  For a single
encounter that dominates the planets' motion in the vertical direction,
such a kick would also cause the two planets' longitudes of ascending node
to be offset by roughly 180$^\circ$, depending on their eccentricities.
Thus, the instantaneous mutual inclination will be close to the sum of the
two planets' inclinations with respect to their initial plane
(Eq.~\ref{eq:inc}).  Although the alignment of the two planets' nodes will
change in time due to secular perturbations, angular momentum conservation
requires that the mutual inclination remain relatively large.

As for the case of inclination with respect to the initial orbital plane,
equal-mass planets yield much larger $\Delta i$ values than planets with
large mass ratios.  However, in this case the lower-mass equal-mass systems
have higher mutual inclinations than the higher-mass cases.  Ten percent of
the unstable planets in the equal-mass systems without disks had
inclinations larger than 35$^\circ$ ({\tt 3J-3J-3J}), 39$^\circ$ ({\tt
J-J-J}), 42$^\circ$ ({\tt S-S-S}), and 49$^\circ$ ({\tt N-N-N}), compared
with 26$^\circ$ for the {\tt Mixed1} simulations.  The larger inclinations
in lower-mass equal-mass planetary systems are caused by the ease with which
the high-mass planets can eject each other.  The higher-mass systems for
which two planets survive have undergone far fewer close encounters than
the lower-mass systems with two surviving planets: the median number of
encounters in the unstable {\tt 3J-3J-3J}, {\tt J-J-J}, {\tt S-S-S}, and
{\tt N-N-N} systems for which two plants survived was 81, 175, 542, and
1871, respectively.  However, the mass-weighted eccentricities of the
two-planet equal-mass systems were slightly higher for the higher-mass
planets.  Thus, large planetary eccentricities are linked with the strength
of encounters between planets while large inclinations are linked with a
large number of scattering events.  Note that the number and strength of
close encounters in systems with significant mass ratios is far less, thus
explaining their lower eccentricities and inclinations (see Raymond \etal
2009a).

As seen in Fig.~\ref{fig:tinst}, the duration of the close encounter phase
(as well as the total number of encounters) is reduced for the simulations
with planetesimal disks.  Therefore, we expect lower inclinations for
simulations with planetesimals -- this is indeed clearly seen in
Fig.~\ref{fig:icum}.  The extreme damping for the {\tt N-N-N} simulations
is again seen in terms of their very low inclinations and mutual
inclinations in systems with planetesimal disks.

\subsection{An angular momentum argument for the transition mass}
What sets the boundary between typically eccentric (and inclined) and
typically circular (and coplanar) planets, which we have determined
empirically lies at a total system mass of about $M_{\rm tot} \approx 0.7 \
M_{\rm J}$?  Plausibly, this transition mass may be set simply by the reservoir
of disk angular momentum that is able to interact with the planet and
circularize its orbit.

To illustrate the point, we construct an (over)-simple model of the 
circularization process. Let us assume that scattering (with or without 
disks) typically yields a single planet with some characteristic 
semi-major axis $a$, eccentricity $e$, and inclination $i$. The 
orbital angular momentum deficit ($AMD$) quantifies the difference 
between the orbital angular momentum for a circular orbit as 
compared to an eccentric and inclined one at the same semi-major 
axis. The $AMD$ is given by,
\begin{equation}
 AMD = M \sqrt{GM_*a} \left( 1 - \cos i \sqrt{1-e^2} \right).
\end{equation}
To fully circularize the planet, we need to add this much angular 
momentum via planetesimal interactions. If these interactions occur 
primarily at or near apocenter in a disk with surface density profile,
\begin{equation}
 \Sigma = C r^{-1},
\end{equation}
with $C$ being a constant, the mass of the disk within a radial zone 
of half-width $n$ Hill radii (at the apocenter distance) is,
\begin{equation}
 \Delta M_{\rm disk} = 4 \pi n C \left( \frac{M}{3 M_*} \right)^{1/3} a (1+e).
\end{equation}
Assuming the planetesimals to have circular orbits, the total angular 
momentum of the disk that interacts with the planet is, approximately,
\begin{equation}
 \Delta L_{\rm disk} \simeq  \Delta M_{\rm disk} \sqrt{GM_* a(1+e)}.
\end{equation}  
By equating the available angular momentum to the angular momentum 
deficit, we find that the disk can circularize planets for masses,
\begin{equation}
 M \lesssim \frac{ (4 \pi n C)^{3/2} }{ (3 M_*)^{1/2} } 
 \frac{ (1+e)^{9/4} }{ (1 - \cos i \sqrt{1-e^2} )^{3/2} } a^{3/2}.
\end{equation}
This is an analog of the usual isolation mass, except here in the case 
of circularization rather than accretion. One should note that the 
factor involving $e$ is {\em not} of order unity -- typically it is 
quite large ($\sim 10^2$).  

This analysis is quite crude, but it does predict transition masses 
that are of the same order of magnitude as those observed. For example, 
substituting our disk parameters and adopting planetary orbital 
elements of $a = 8 \ {\rm AU}$, $e = 0.4$ and $i = 10^\circ$, we 
obtain (for $n=2$) a transition mass $M \approx 0.6 \ M_{\rm J}$. This 
suggests that we should interpret the transition mass as being a 
consequence of the finite angular momentum, within the disk, available 
to circularize planets. So, why is it that the eccentricity is
controlled by the total planet mass rather than the individual planet mass
(e.g., Figs.~\ref{fig:e-mtot} and~\ref{fig:e-m})?  The answer appears to be
that it is the total mass which regulates the eccentricity excitation,
which in turn determines the amount of damping by scattering disk
particles.

\subsection{Radial Mass Distributions}

Planet-planet scattering tends to segregate systems by mass, with more
massive planets closer-in and less massive planets farther out (Chatterjee
\etal 2008).  We expect this effect to be magnified in systems with
planetesimal disks because the disk can ``trap'' low-mass planets that
would otherwise be ejected from the system.  To look at the radial mass
evolution of our simulations we restrict ourselves to the {\tt Mixed1} and
{\tt Mixed2} cases which did not start with pre-defined radial mass
gradients.

\begin{figure}
\centerline{\plotone{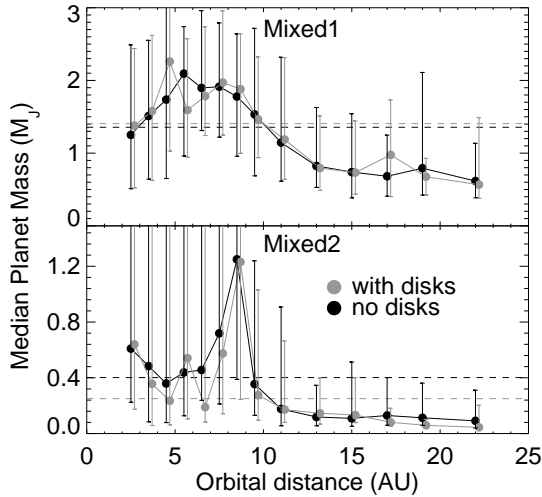}}
\caption{Radial mass distribution of scattered planets for the {\tt Mixed1}
and {\tt Mixed2} simulations, with (grey) and without (black) planetesimal
disks.  The dots represent the median planet mass in a given radial bin,
and the error bars show the 10\% to 90\% range in values.  The point at 22
AU includes all planets beyond 20 AU.  The dashed horizontal lines show the
median planet masses for the ensemble of all scattered planets in a given
set.  Note that the simulations with disks are offset by 0.2 AU for
clarity.}
\label{fig:mmed-a}
\end{figure}

Figure~\ref{fig:mmed-a} shows the median planet mass as a function of
orbital distance for the unstable {\tt Mixed1} and {\tt Mixed2} systems,
with and without disks (also shown is the 10-90\% range of planet masses in
a given radial bin).  For the {\tt Mixed1} systems, there is a clear
segregation of high-mass planets to the inner system and low-mass planets
in the outer system, with very little difference for the simulations with
and without planetesimal disks.  The most massive planets tend to reside at
4-9 AU, close to the starting initial orbits.  

Radial mass segregation is also clear in the {\tt Mixed2} simulations but
there are additional subtleties.  As for the {\tt
Mixed1} cases, high-mass planets are confined within 10 AU and low-mass
planets outside 10 AU. There is a large peak in the median
{\tt Mixed2} planet mass between 8 and 9 AU both with and without
planetesimal disks. There are actually 2-3 times more planets in this 1
AU-wide bin than in adjacent bins for the {\tt Mixed2} simulations with and
without disks. Perhaps surprisingly, any differences in mass segregation  
caused by disks appear to be modest even for the {\tt Mixed2} simulations.

\begin{figure}
\centerline{\plotone{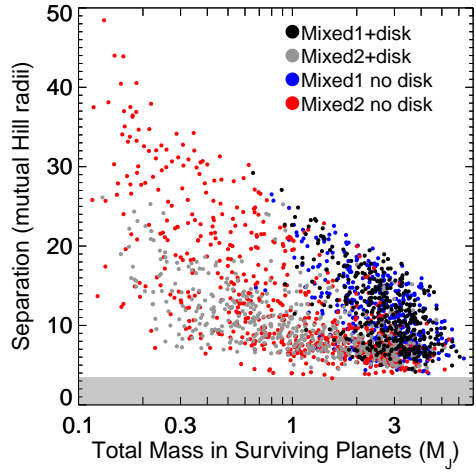}}
\caption{Final planetary separation in units of mutual Hill radii for
  unstable {\tt Mixed1} and {\tt Mixed2} simulations, with (black and
grey)
  and without (blue and red) planetesimal disks.  Only systems with two
or
  more surviving planets are shown -- for systems with three surviving
  planets only the separation between the inner two planets is included.
  The shaded horizontal line shows the two-planet stability limit of $3.46 R_{H,m}$ 
(Marchal \& Bozis 1982; Gladman 1993). }
\label{fig:mhr-mtot}
\end{figure}

Scattering tends to spread planetary systems out. We can quantify this by
looking at the planetary separations at the end of the simulations in units
of mutual Hill radii $R_{H,m}$ (recall that all simulations started
separated by 4-5 $R_{H,m}$). Figure~\ref{fig:mhr-mtot} shows the separation
of the two inner planets as a function of the total mass in surviving
planets for the {\tt Mixed1} and {\tt Mixed2} simulations, with and without
disks. There are several interesting pieces of information in
Fig.~\ref{fig:mhr-mtot}.  First, the typical interplanetary spacing is
$\sim$ 4-30 $R_{H,m}$, showing that most systems have indeed spread
out. Second, the interplanetary spacing decreases significantly for larger
system masses, although there remains a large spread for any given
mass. This general trend can be explained by the much larger number of
scattering events undergone by lower-mass systems before the destruction
(usually by ejection) of a planet\footnote{The separations in
Fig.~\ref{fig:mhr-mtot} were calculated using the orbital semimajor axes. A
more important criterion in terms of dynamical stability is the closest
approach distance between the two planets, i.e., the difference between the
outer planet's perihelion distance and the inner planet's aphelion. We have
looked at that distribution and the negative slope is much less steep than
in Fig~\ref{fig:mhr-mtot}, indicating that eccentricity plays a significant
role.}.  The spread in separation is due in part to stochastic variations in
scattering events from simulation to simulation and in part to variations
in planetary mass ratios -- the larger number of scattering events for
equal-mass systems causes them to be more widely-spaced than systems with
larger mass ratios. Third, there is an abrupt break between the separations
of lower-mass systems with and without planetesimal disks. For $M_{tot}
\lesssim 0.7 M_{\rm J}$, systems with disks are much more compact than systems
without disks. This is a result of the damping of the planetesimal disk:
low-mass planets on wide orbits have their eccentricities efficiently
damped and their perihleion distances increased such that they avoid
additional close encounters with the inner planet. This is evidenced by the
fact that the break between the simulations with and without disks occurs
at the same total planet mass as for the eccentricity and inclination
distributions (see Figs~\ref{fig:e-mtot} and~\ref{fig:mtot-iin}).

\begin{figure}
\centerline{\plotone{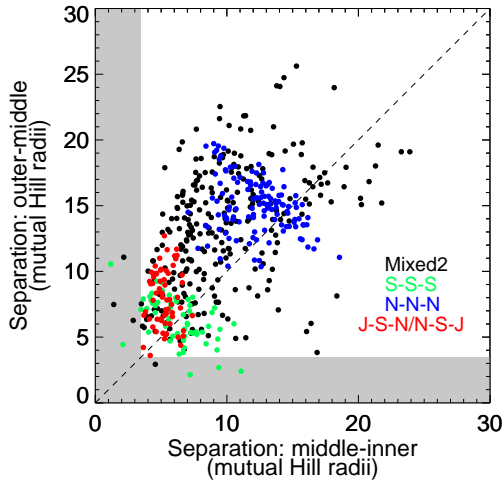}}
\caption{For scattered systems with three surviving planets, the
separation
  of the outer planet pair ($y$ axis) vs. the inner planet pair ($x$
axis),
  in units of mutual Hill radii.  Five sets of simulations are shown in
  different colors.  The shaded horizontal and vertical lines shows the
  two-planet stability limit of 3.46 $R_{H,m}$. }
\label{fig:mhrin-out}
\end{figure}

For systems in which three planets survive, the outer two planets tend to
be more widely spaced than the inner two. Figure~\ref{fig:mhrin-out} shows
the separation of each pair of planets for unstable systems which preserved
all three planets. In the bulk of cases the outer pair of planets are
indeed more widely-spaced than the inner pair. However, there do exist many
cases for which the outer planets are more compact, as well as a few cases
in the Hill unstable region which are likely to be unstable on
slightly longer timescales. Systems with a more compact outer pair of
planets tend to be those with lower-mass {\em middle} planets. As the outer
planet scatters planetesimals inward, a massive middle planet can eject
them from the system, causing the planet to move inward and leading to a
compact inner system but a spread out outer system. In contrast, a low mass
middle planet cannot generally eject the scattered planetesimals and so
simply scatters them toward the inner planet, causing the middle planet to
move outward, yielding a compact outer system and a more spread out inner
system.

\subsection{Planets at large orbital separations}
Planet-planet scattering creates a population of high-eccentricity planets
with large apocenter distances (Veras \etal 2009; Scharf \& Menou 2009).
For the vast majority of cases, these planets, which can have semimajor
axes as large as $\gtrsim 10,000$ AU, represent a transient phase on the
path to dynamical ejection\footnote{In a few cases the planets can be
stabilized by external torques such as from the potential of the galactic
disk (Veras \etal 2009).}.  These planets are of interest because they
represent the source of so-called ``free-floating planets''. In some
cases their large separations mean that they could be good targets 
for direct detection, although they probably only exist for the first 10-100
Myr of a star or star cluster's lifetime.

\begin{figure*}
\centerline{\plotone{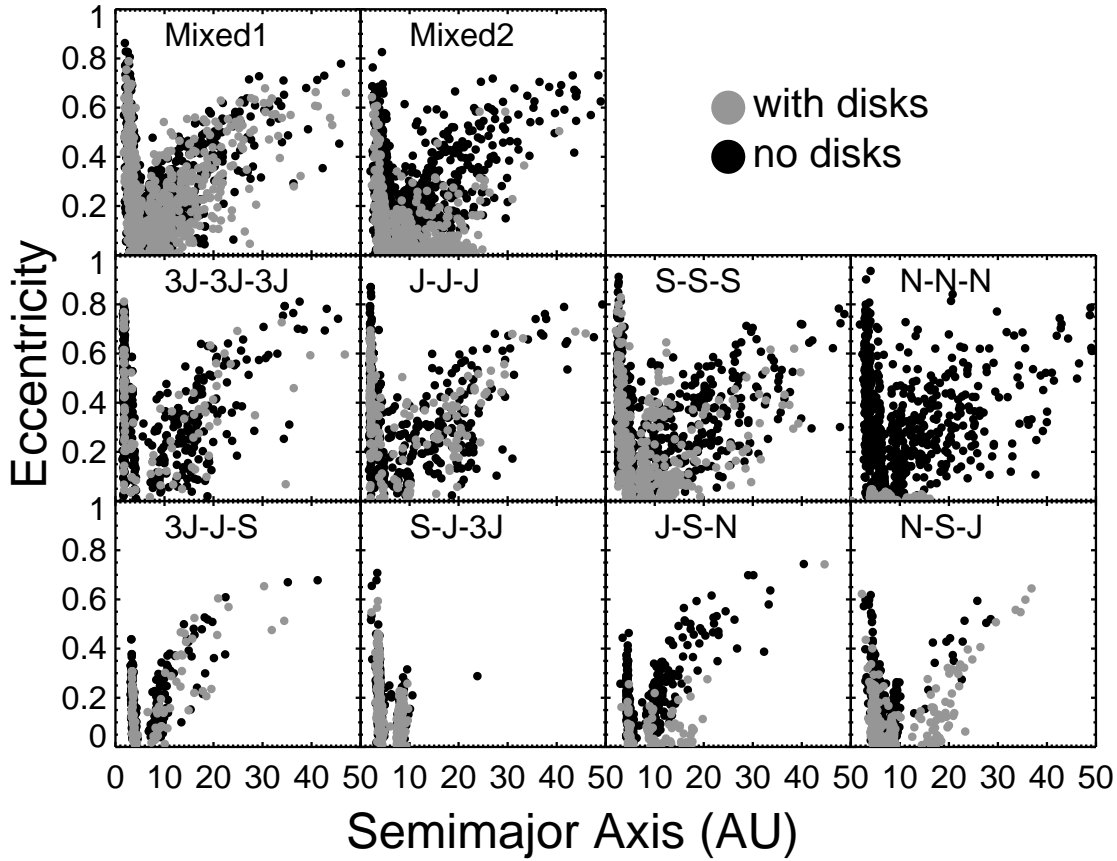}}
\caption{Orbital eccentricity vs. semimajor axis of scattered systems with
(in grey) and without (in black) planetesimal disks, for each of our ten
initial mass distributions. }
\label{fig:a-e_all}
\end{figure*}

In our simulations we imposed an ejection radius of 100 AU, so we were
unable to probe very distant planets. Nonetheless, we can study planets
that ended up on transient, or in some cases stable, orbits beyond the
initial radial extent of our initial conditions. To address the issue of
planets on widely-separated orbits, we kept track of the time spent by each
planet in each simulation with an aphelion distance in radial bins of 25-50
AU, 50-75 AU, and 75-100 AU. We also characterized planets in each of those
bins as either being in a transitional state (usually on the path to
ejection) or on stable orbits. Inspection of the final eccentricity as 
a function of semi-major axis distributions, plotted in Fig.~\ref{fig:a-e_all}, 
shows immediately that a large number of scattered planets
survived on stable orbits with large aphelion distances.

\begin{table*}
\center
\caption{Scattered planets at large orbital distances (aphelion
$Q>25[50]$ AU)$^\star$}
\begin{tabular}{|l|ccc|ccc|}
\hline
& \multicolumn{3}{c|}{No Disks} & \multicolumn{3}{c|}{With Disks} \\
\cline{2-4}
\cline{5-7}
Set & frac stable & frac trans. & $\langle t \rangle$ (Myr) & frac stable & frac
trans. & $\langle t \rangle$ (Myr) \\
\hline
{\tt Mixed1} & 0.14 [0.03] & 0.44 & 0.64 [0.10] & 0.17 [0.03] & 0.45 & 0.92 [0.15] \\
{\tt Mixed2} & 0.15 [0.03] & 0.59 & 2.55 [0.54] & 0.03 [0.00] & 0.23 & 0.46 [0.03] \\
{\tt 3J-3J-3J} & 0.23 [0.05] & 0.23 & 0.69 [0.01] & 0.29 [0.07] & 0.29 & 2.60 [0.01] \\
{\tt J-J-J} & 0.29 [0.06] & 0.49 & 0.77 [0.03] & 0.29 [0.04] & 0.53 & 1.88 [0.02] \\
{\tt S-S-S} & 0.34 [0.07] & 0.64 & 1.36 [0.28] & 0.18 [0.04] & 0.46 & 2.19 [0.28] \\
{\tt N-N-N} & 0.32 [0.08] & 0.63 & 7.58 [1.69] & 0.00 [0.00] & 0.00 & 0.00 [0.00] \\
{\tt 3J-J-S} & 0.06 [0.01] & 0.31 & 0.67 [0.08] &  0.11 [0.04] & 0.29 & 0.19 [0.01] \\
{\tt S-J-3J} & 0.00 [0.00] & 0.14 & 0.01 [0.01] & 0.00 [0.00] & 0.11 & 0.01 [0.00] \\
{\tt J-S-N} & 0.13 [0.02] & 0.59 & 1.99 [0.33] & 0.05 [0.02] & 0.33 & 2.45 [0.66] \\
{\tt N-S-J} & 0.05 [0.00] & 0.59 & 0.16 [0.06] & 0.12 [0.03] & 0.41 & 1.30 [0.14] \\

\hline
\end{tabular}
\\
$^\star$ The first column (``frac stable'') represents the fraction of
scattered systems containing a planet on a long-term stable orbit with
aphelion distance $Q$ larger than 25 [50] AU. The second column (``frac
trans'') shows the fraction of scattered systems for which at least one
planet spent at least one output interval of $10^5$ years with $Q > 25$
AU
during a transitional ejection phase. The third column shows the mean
time
$\langle t \rangle$ in Myr during which a planet had $Q >$ 25 [50] AU for the
transitional systems. Note that $\langle t \rangle$ is averaged over all scattered
systems, including a fraction that did not undergo a transitional outer
planet phase.
\end{table*}

Table 2 shows the statistics of planets at large orbital separations in our
simulations. The table includes information about both stable and
transitional planets with aphelion distances $Q > 25$~AU and $Q > 50$~AU. First of
all, we see that scattering among equal-mass planets is far more efficient
at producing stable planets with large orbital radii than scattering among
planets with mass gradients. This is simply because of the much larger
number of scattering events that occur in equal-mass systems. Next, we see
that this trend does not hold for the fraction of systems which experience
a transitional, high-$Q$ phase. Lower-mass planets have a higher
probability of experiencing this transitional phase. We interpret this as
an ejection timescale issue -- the number of encounters and duration of the
instability phase increases dramatically for lower-mass systems (see
Section 3.1) such that the typical ejection is very drawn out. In contrast,
for higher-mass planets a smaller number of scattering events is needed to
eject a planet. Our analysis only includes outputs every $10^5$ years;
thus, ejections which occur in less than that time are not registered or
counted in Table 2. For equal-mass and mixed systems, the mean time during
our 100 Myr simulations for which a planet could be observed with $Q >$ 25
[50] AU correlates with the fraction of transitional systems, with typical
observable probabilities of a couple of percent (the probability is here 
calculated as $\langle t
\rangle$ divided by the 100 Myr simulation length). For the mass gradient
simulations, $\langle t \rangle$ is longer for negative mass gradients (for
{\tt 3J-J-S} and {\tt J-S-N}). This is because the most massive planet is
closer-in in those cases, deeper in the star's potential well, such that a
larger number of encounters is needed to reach
the system's escape velocity.

As expected, the higher-mass systems are barely affected by the presence of
planetesimals disks in terms of the fraction of orbits which are at large
distances and the fraction that appear in the transitional phase.  However,
the duration of the transitional ejection phase is lengthened by a factor
of a few for the equal-mass cases, meaning that the time required to eject
those planets is increased due to the damping from the disk. For the
lower-mass planets there are fewer planets at large orbital distances in
terms of both stable and transitional configurations. This is simply due to
the disk's ability to ``trap'' planets by orbital circularization. This
circularization occurs within the disk, so trapped planets are generally
confined to orbits within the outer edge of the disk, at 20 AU.

Can observations of long-period planets tell us anything about the inner
planets that spawned them? Assuming that outer planetesimal disks are
ubiquitous, equal-mass and higher-mass planetary systems are more likely to
populate distant stable orbits than lower-mass systems or those with large
mass ratios between planets. However, a wide range of configurations
produces long-lived transient planets on large orbits that are generally on
their way towards dynamical ejection. We suspect that combining observations of
long-period planets with infrared observations of the outer
planetesimal disk may constrain the problem, but that is beyond the scope
of the current paper.

\subsection{Planetary System Packing}

A more dynamically sophisticated way to interpret the radial distribution
of planets in multiple planet systems is to measure their separation in
terms of the minimum required for Hill stability. This approach
incorporates eccentricity and inclination information consistently and, if
there are {\em only} two planets in the system, connects to the formal
definition of Hill stability that has been proven for that case (Marchal \&
Bozis 1982; Gladman 1993).  Observationally, the known two-planet
exosystems are known to cluster close to the Hill stability boundary
(Barnes \& Greenberg 2006). This motivates the hypothesis that {\em all}
multiple planet systems may be dynamically ``packed", in the sense that
additional planets could not exist between the known planets on
stable orbits (see Barnes \etal 2008 or Raymond \etal 2008b). The corollary
of such a ``packed planetary system" hypothesis (Barnes \& Raymond 2004;
Raymond \& Barnes 2005; Raymond \etal 2006; see also Laskar 1997), of
course, is that when planetary systems are observed {\em not} to be packed
we should seek an additional ``missing" (normally lower mass) planet in
what appears to be an empty stable orbit.

It is easy to test numerically whether an observed multiple planet system
is packed (e.g. Rivera \& Lissauer 2000; Jones \etal 2001; Menou \&
Tabachnik 2003; Asghari \etal 2004; Raymond \etal 2008b; Kopparapu \etal
2009). In one interesting case, the HD 74156 system was dynamically mapped
to reveal a narrow zone between the two known planets that was stable for
Saturn-mass test planets but not for Jupiter-mass test planets (Raymond \&
Barnes 2005). Subsequently, Bean \etal (2008) presented evidence for a 
third planet in the system, in the stable zone and with a mass of 1.4 
Saturn masses (Bean \etal 2008; Barnes \etal 2008)\footnote{Note that a 
recent analysis of Hobby-Eberly Telescope data by Wittenmyer \etal (2009) failed 
to confirm the presence of HD~74156~d, hence the current status of this 
planet is unclear.}. In a previous paper we showed that
pure planet-planet scattering naturally reproduces the observed
distribution of dynamical configurations (Raymond \etal 2009a).  Given the
significant effects that planetesimal disks can have on planetary
evolution, here we investigate whether this is still true at radii where
the dynamical influence of disks is significant.

For two planets with masses $M_1$ and $M_2$ orbiting a star, it can be
shown analytically that long-term dynamical stability requires,
\begin{eqnarray}
\frac{-2 (M_*+M_1+M_2)}{G^2 (M_1 M_2 + M_* M_1 + M_* M_2)^3} \,
c^2 h \geq \nonumber \\ 1 + 3^{4/3} \frac{M_1 M_2}{M_*^{2/3} (M_1+M_2)^{4/3}} -
\frac{M_1 M_2 (11 M_1 + 7 M_2)}{3 M_* (M_1 +M_2)^2}, 
\label{eqn:beta1}
\end{eqnarray}
\noindent where $c$ and $h$ represent the total orbital angular momentum and
energy of the system, respectively (Marchal \& Bozis 1982; Gladman 1993; Veras
\& Armitage 2004; note that this definition assumes that $M_1 > M_2$).  
Following Barnes \& Greenberg (2006), we refer to the left side of
Eqn~\ref{eqn:beta1} as $\beta$ and the right side as $\beta_{crit}$.  The
quantity $\bbc$ therefore measures the proximity of a pair of orbits to the
Hill stability limit of $\bbc=1$.  It is important to note that our $\bbc$
analysis only applies for two-planet systems, because perturbations from
additional companions can shift the stability boundary to values other than
1.

In Raymond \etal (2009a) we calculated $\bbc$ by ``observing'' each system
from a wide range of viewing angles, then using Eqn~\ref{eqn:beta1} to
calculate $\bbc$ distributions.  Although the $\bbc$ value is sensitive to
both the inclination between the planetary orbital plane and the line of
sight, $I$, and the mutual inclination between the planets' orbits, $\Delta
i$, we found that the effect of these angles was at the few percent level
at most, i.e., negligible for our purposes.  Additional tests have showed
that a simple application of Eqn~\ref{eqn:beta1} with no knowledge of
viewing angle agrees remarkably well with the true and observed $\bbc$
values as well as the cumulative distribution from a range of viewing
angles.  Thus, in this analysis we do not include the effects of viewing
geometry.

\begin{figure*}
\centerline{\plotone{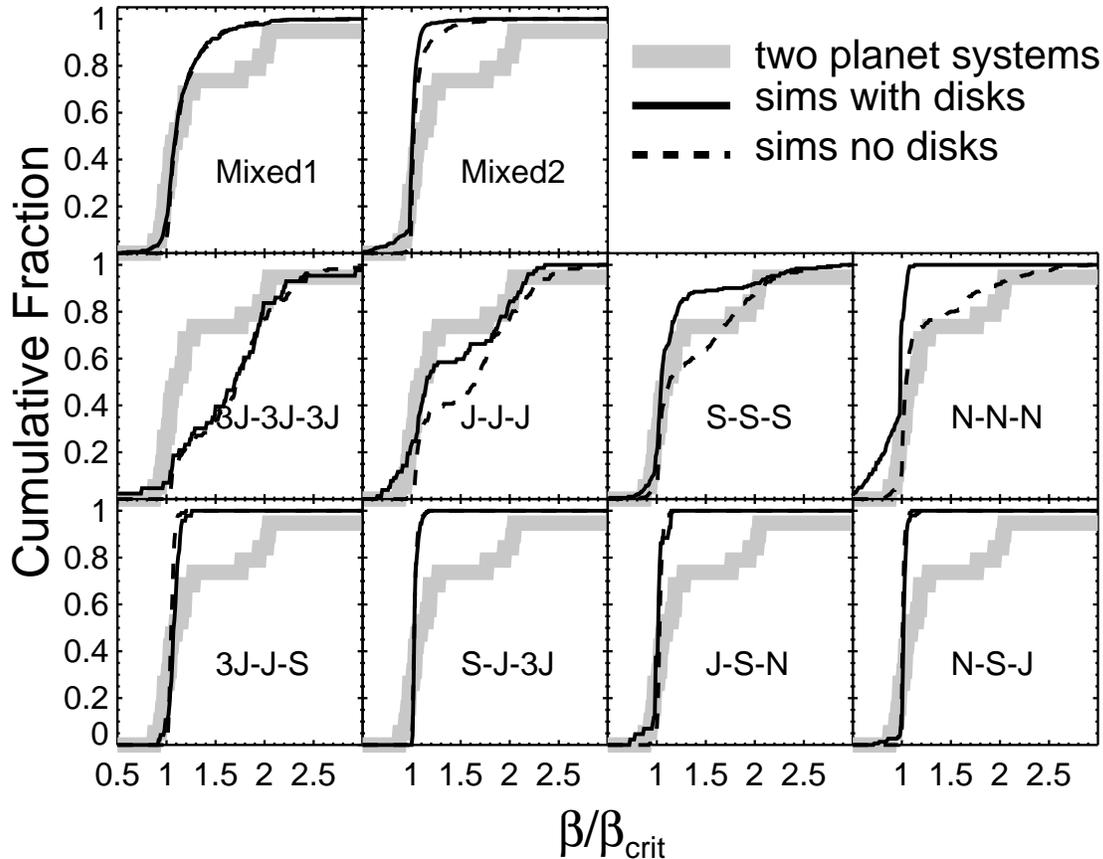}}
\caption{Cumulative $\bbc$ distributions for the unstable simulations
from each of our ten simulation cases, with (solid lines) and without
(dashed lines) planetesimal disks, compared with the observed extra-solar
systems exterior to 0.1 AU (thick grey line).  Note that one system,
HD47186 (Bouchy \etal 2009), has $\bbc \sim 6$.}
\label{fig:betacum_all}
\end{figure*}  

Figure~\ref{fig:betacum_all} compares the cumulative $\bbc$ distributions
for our unstable simulations with the observed two planet extra-solar
systems\footnote{Data taken from
http://www.astro.washington.edu/users/ rory/research/xsp/dynamics/ on June
24, 2009.  We excluded three systems for which the inner planet was
interior to 0.1 AU and therefore had probably undergone tidal evolution,
thereby increasing the effective separation between the planets and
therefore the $\bbc$ value.  Our total exoplanet sample includes 19
systems.  For the simulations without disks we only applied the analysis to
systems with two surviving planets, but for the simulations with disks we
included unstable systems with two or three surviving planets; for the case
of three surviving planets we use the $\bbc$ value calculated by assuming
detection of only the two inner planets.}.  As seen in Raymond\etal
(2009a), $\bbc$ values from systems with equal-mass planets are far larger
than those from systems with planetary mass ratios.  This appears to be
caused by the vastly larger number of encounters among equal-mass planets
before a planet is destroyed.

For the simulations with the most realistic mass distributions (i.e. the
{\tt Mixed1} and {\tt Mixed2} cases) the effect of disks on the $\beta /
\beta_{\rm crit}$ is small. Disks do not alter the prediction that
scattering yields packed systems. Where there are differences they are in a
surprising direction -- after an instability systems with planetesimal
disks remain {\em closer} to the stability boundary (i.e., have lower
$\bbc$ values) than systems with no planetesimal disks.  Since planetesimal
disks damp eccentricities, they appear to leave the surviving planets in
more dynamically compact configurations.  This is true even for the
low-mass systems such as {\tt N-N-N}, for which planetesimal scattering
almost always induces the migration of one planet out into disk.  However,
although the outer planet often migrates outward in {\tt N-N-N}-type cases,
the orbits of the inner planets are often compressed by scattering
planetesimals outward.  So, if we just consider the inner two planets in
any system, they appear to end up closer together in the cases with
planetesimal disks.

\begin{table*}
\center
\caption{$p$ values from K-S tests of observations vs. scattering
simulations$^\star$} 
\begin{tabular}{|l|ccc|ccc|}
\hline
& \multicolumn{3}{c|}{No Disks} & \multicolumn{3}{c|}{With Disks} \\
\cline{2-4}
\cline{5-7}
Set & $p$ & $p$ & $p$ & $p$ & $p$ & $p$ \\
    & (tot) & $(\bbc \leq 1)$ & $(\bbc > 1)$ & (tot) & $(\bbc \leq 1)$ &
    $(\bbc > 1)$ \\
\hline
   {\tt Mixed1} & 0.175 & 0.005 & 0.074  &  0.226 & 0.818 & 0.091 \\  
   {\tt Mixed2} & 0.015 & --- & 0.005  	 &  --- & 0.005 & --- \\ 	   
 {\tt 3J-3J-3J} & --- & --- & 0.016    	 &  --- & 0.012 & 0.010 \\    	   
    {\tt J-J-J} & 0.015 & 0.012 & 0.195	 &  0.470 & 0.111 & 0.378 \\  
    {\tt S-S-S} & 0.292 & 0.425 & 0.936	 &  0.476 & 0.328 & 0.215 \\  
    {\tt N-N-N} & 0.439 & 0.438 & 0.284	 &  --- & 0.089 & --- \\ 	   
   {\tt 3J-J-S} & --- & --- & ---    	 &  0.002 & 0.313 & --- \\    
   {\tt S-J-3J} & --- & 0.012 & ---    	 &  --- & 1.000 & --- \\ 	   
    {\tt J-S-N} & --- & 0.007 & ---    	 &  --- & 0.008 & --- \\ 	   
    {\tt N-S-J} & --- & --- & ---        &  --- & 0.183 & --- \\      
\hline
\end{tabular}
\\ 
$^\star$The majority of $p$ values of '---' indicate $p < 10^{-3}$, and in
a few cases insufficient data (e.g., none of the {\tt 3J-3J-3J} simulations
had $\bbc \leq 1$).
\end{table*}

Without disks, the {\tt Mixed1}, {\tt S-S-S} and {\tt N-N-N} simulations
all provide good fits to the $\bbc$ distribution -- Table 3 contains $p$
values from K-S tests for each case.  However, none of the sets provides a
good match for $\bbc \leq 1$: the {\tt S-S-S} and {\tt N-N-N} cases are
contaminated by systems which are themselves unstable on timescales longer
than our 100 Myr integration time (these were removed by hand in Raymond
\etal 2009a but not here).  With disks, the {\tt S-S-S} simulations provide
the best match to the known systems and the {\tt Mixed1} and {\tt J-J-J}
simulations also provide good fits.  In addition, several sets match the
distribution very well for $\bbc \leq 1$ with disks while they failed
without disks.  The cause of this increase in low-$\bbc$ systems is the
presence of mean motion resonances.  Systems in resonance tend to have
$\bbc < 1$ (Barnes \& Greenberg 2007), and these are preferentially set up
in systems with planet masses comparable to $M_{\rm J}$ such as the {\tt Mixed1}
and {\tt J-J-J} simulations (see \S 4 below).

\section{Mean Motion Resonances (MMRs)}

A significant fraction of the known exoplanet systems are thought to lie in
mean motion resonances (henceforth MMRs). Marcy et al. (2008), for example, 
quote a resonant fraction of 18\% (4 out of 22 systems). These numbers are 
currently uncertain because confirmation that systems near resonance are 
actually librating is often hard to obtain -- indeed apparently resonant
systems are sometimes revised to be non-resonant (e.g., Fischer \etal
2008). The most likely origin of resonant planets at small orbital radii is 
convergent migration in gaseous protoplanetary disks (Snellgrove \etal
2001; Lee \& Peale 2002; Kley \etal 2004). If this identification is correct, 
it places bounds on the strength of turbulence within the disk, since 
even rather modest fluctuations in the disk surface density should  
destroy resonant alignment (Adams \etal 2008; Lecoanet \etal 2009).  
Here, we investigate gas free routes to the establishment of resonance due 
to either pure planet-planet scattering or scattering augmented by the 
dynamical effect of planetesimals. These channels might contribute to 
the population of resonance systems at small radii, and potentially 
dominate it further out where the effects of planetesimal disks are strong.

In previous work related to planet-planet scattering, we studied two 
mechanisms that can lead to MMRs.  First, scattering in isolation (without
disks) populates a variety of resonances including high-order MMRs (up to
11th order in our simulations; Raymond \etal 2008a).  Second, the
planetesimal disk can effectively act as a damping force on planets' orbits
to induce planetary orbits to align into MMRs and sometimes MMR chains
(e.g., 4:2:1) with a high efficiency (Raymond \etal 2009b).  In this
section we study MMRs that arose in both our unstable simulations (via
scattering) and our stable simulations (via planet-planetesimal effects).
We first introduce the theory of MMRs in \S 4.1, then examine resonances in
unstable (\S 4.2) and stable (\S 4.3) systems.

\subsection{Definition of Mean Motion Resonances}

For mean motion resonance $p+q:p$, resonant arguments $\theta_i$ (also
called ``resonant angles'') are of the form
\begin{equation}
\theta_{1,2} = (p+q) \lambda_1 - p \lambda_2 -q \varpi_{1,2} \\
\label{eqn:mmr}
\end{equation}
\noindent where $\lambda$ are mean longitudes, $\varpi$ are longitudes
of pericenter, and subscripts 1 and 2 refer to the inner and outer planet,
respectively (e.g., Murray \& Dermott 2000).  Resonant arguments measure
the angle between the two planets at the conjunction point -- if any
argument librates rather than circulates, then the planets are in
resonance.  In fact, the bulk of resonant configurations are characterized
by only one librating resonant argument (Michtchenko \etal 2008). Frequently, 
libration occurs around equilibrium angles of zero or 180$^\circ$, 
but any angle can serve as the equilibrium.  Different resonances have
different numbers of resonant arguments, involving various permutations
of the final terms in Eq.~\ref{eqn:mmr}.  The order of a resonance is given
simply by $q$.  For example, the 3:1 MMR ($q=2, p=1$) is a second-order
resonance that has three resonant arguments:
\begin{eqnarray}
\theta_{1} &=& 3 \lambda_1 - \lambda_2 - 2 \varpi_{1}, \hskip .2in
\nonumber \\
\theta_{2} &=& 3 \lambda_1 - \lambda_2 - 2 \varpi_{2}, \hskip .1in {\rm and } \hskip .2in
\nonumber \\
\theta_{3} &=& 3 \lambda_1 - \lambda_2 -  (\varpi_{1}+\varpi_2).
\end{eqnarray}

In Raymond \etal (2008a), we found MMRs by targeting systems close to
commensurabilities, then looking at individual resonant angles.  In this
paper we take an automated approach.  We devised a simple analysis script
to analyze each of our simulations (with and without disks) and calculate
all resonant angles for MMRs of up to fourth order.  This includes the 2:1,
3:2, 3:1, 5:3, 5:2, and 4:1 MMRs.  The script then determined which resonant 
angles were librating during the last 50
Myr of each simulation.  This script is simple and reliable, but it misses
the small fraction of cases that enter resonance between 50 and 100 Myr because
it requires $\theta$ values to librate for the full interval (50-100 Myr).
We impose a cutoff on the libration amplitude $A$ of $A \leq
150^\circ$ to avoid false positives, which eliminates some additional
systems that are just barely resonant.  

\subsection{MMRs in unstable systems}

In previous work, we found that a few to ten percent of unstable systems
could end in MMRs (Raymond \etal 2008a).  The resonant fraction depends on
the initial planetary mass distribution: MMRs were much more common in
systems with radial mass gradients than in simulations that started with
equal-mass planets.  MMRs are populated by scattering simply because the
density of resonant orbits is non-zero.  In other words, the last close
encounter leading to the destruction of one planet deposits the surviving
planets onto orbits which have a chance of being resonant.  This mechanism
can populate high-order MMRs that are not populated by convergent migration
in gaseous disks (Snellgrove \etal 2001; Lee \& Peale 2002).  The libration
amplitudes of resonances populated by scattering are large, which also
contrasts with the convergent migration scenario.  Thus, as the population of
resonant exoplanet systems increases, we may be able to differentiate MMRs
caused by scattering vs. convergent migration. 

\begin{figure}
\epsscale{1.2}
\centerline{\plotone{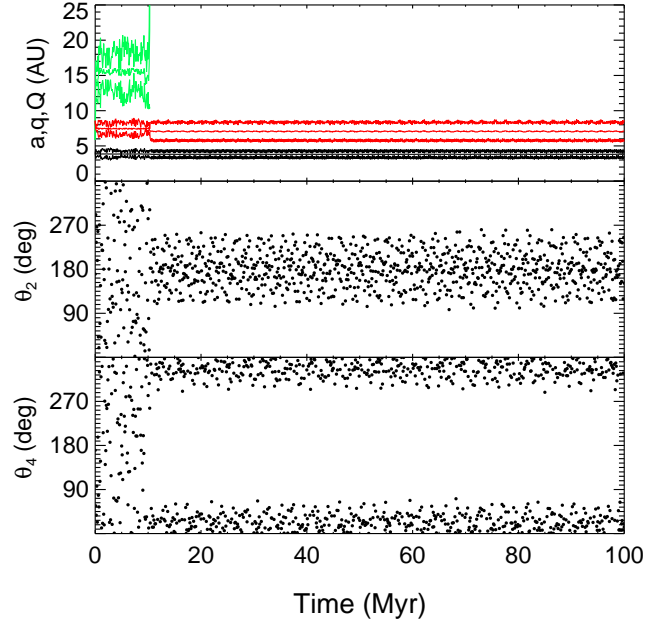}}
\caption{A 5:2 mean motion resonance set up by scattering.  The top panel
shows the evolution of the three planets' semimajor axes $a$, perihelion
distances $q$ and aphelion distances $Q$.  The middle and bottom panel show
two resonant angles for the 5:2 MMR: $\theta_2 = 5\,\lambda_1 - 2
\,\lambda_2 - 3 \, \varpi_2$, and $\theta_4 = 5\,\lambda_1 - 2
\,\lambda_2 - 2 \, \varpi_2 - \varpi_1$.}
\label{fig:mmr_scat}
\epsscale{1.0}
\end{figure}

Figure~\ref{fig:mmr_scat} shows an example of an MMR caused by
planet-planet scattering in a simulation that included a planetesimal disk.
After 30,000 years, a close encounter with the outer (1.8 $M_{\rm J}$) planet
threw the lower-mass (0.4 $M_{\rm J}$) planet outward into the planetesimal disk,
where it remained on a stable but chaotic orbit for about 10 Myr.  A final
close encounter caused the lower-mass planet to be ejected and placed the
surviving (1.9 and 1.8 $M_{\rm J}$) planets in 5:2 MMR.  In fact, the two
surviving planets are in a special type of resonance called an apsidal
corotation resonance in which all resonant angles librate and so does the
apsidal alignment (Beaug{\'e} \etal 2003). This simulation was unusual because the entire planetesimal disk
had been destroyed by the outer planet by the time of the last planetary
scattering event such that the behavior was similar to simulations without
disks.

Most MMRs that occur in unstable systems with disks arise via a different
mechanism.  Usually, an instability occurs early in the simulation,
removing one planet and placing the two surviving massive ($M_p \sim M_{\rm J}$)
planets close to resonance.  Subsequent interactions with the planetesimal
disk act as a damping force that aligns the planetary orbits into MMR.  In
these cases the outer surviving planet is massive enough to eject most of
the planetesimals that it encounters, leading to some inward migration and
a compression of the system.  In fact, this mechanism has more in common
with the generation of MMRs and MMR chains via planetesimal disk
interactions (see \S 4.3 below) than MMRs from pure N-body scattering.

\begin{table*}
\center
\caption{Mean Motion Resonances in Scattered Systems$^\star$}
\begin{tabular}{|l|ccc|ccc|}
\hline
& \multicolumn{3}{c|}{No Disks} & \multicolumn{3}{c|}{With Disks} \\
\cline{2-4}
\cline{5-7}
Set & 1st Order  & 2nd Order  & 3rd Order  & 1st Order
 & 2nd Order  & 3rd Order  \\
    & (frac) & (frac) & (frac) & (frac) & (frac) & (frac)\\
\hline
   {\tt Mixed1} &  10 (0.018) &  9 (0.016) &  3 (0.005) &   5 (0.011) &  2 (0.005) &  1 (0.002) \\
   {\tt Mixed2} &  25 (0.034) & 13 (0.017) &  8 (0.011) &   9 (0.017) &  2 (0.004) &  1 (0.002) \\
 {\tt 3J-3J-3J} &   3 (0.012) &  2 (0.008) &  0 (---) &   0 (---) &  0 (---) &  0 (---) \\
    {\tt J-J-J} &   1 (0.004) &  0 (---) &  2 (0.009) &   0 (---) &  0 (---) &  0 (---) \\
    {\tt S-S-S} &   8 (0.022) &  3 (0.008) &  4 (0.011) &   1 (0.005) &  1 (0.005) &  0 (---) \\
    {\tt N-N-N} &  10 (0.028) &  4 (0.011) &  4 (0.011) &   0 (---) &  0 (---) &  0 (---) \\
   {\tt 3J-J-S} &   1 (0.007) &  1 (0.007) &  2 (0.013) &   1 (0.018) &  0 (---) &  0 (---) \\
   {\tt S-J-3J} &   1 (0.005) & 13 (0.059) &  2 (0.009) &   0 (---) &  6 (0.041) &  2 (0.014) \\
    {\tt J-S-N} &   2 (0.010) &  4 (0.019) &  4 (0.019) &   0 (---) &  0 (---) &  0 (---) \\
    {\tt N-S-J} &  16 (0.072) &  4 (0.018) &  3 (0.014) &   8 (0.054) &  1 (0.007) &  0 (---) \\
\hline
\label{tab:mmr-scat}
\end{tabular}
\\ 
$^\star$The resonance order is given by $q$ in Eqn~\ref{eqn:mmr}.  The
first-order MMRs are 2:1 and 3:2, second-order are 3:1 and 5:3, and
third-order are 4:1 and 5:2.  The values in parentheses represent the
fraction of {\em unstable} simulations that ended up in each set of MMRs.
\end{table*}

The frequency of MMRs from scattering is greatly reduced by the presence of
the planetesimal disk (Table~\ref{tab:mmr-scat}).  In fact, only a small
fraction of the MMRs with disks from Table~\ref{tab:mmr-scat} are even due
to pure scattering -- the majority are caused by disk damping as described
above.  The reason for the absence of MMRs in unstable systems with
planetesimal disks is that the low-order MMRs populated by scattering occur
preferentially in systems with high mass ratios (for example, almost 10\%
of the unstable {\tt N-S-J} simulations yielded 2:1 MMRs), meaning that one
planet in the system was relatively low-mass (usually $\sim M_{\rm S}$) and
subject to radial migration by planetesimal scattering after the
instability placed the planets in resonance.

Thus, scattering appears to be a viable mechanism to populate resonances
but its efficiency is reduced in regions which are strongly affected by
planetesimal disks.  Given that the observed exoplanets do not yet show an
obvious signature of planetesimal scattering (see \S 5 below), some of the
known systems could indeed have been populated by this mechanism (Raymond
\etal 2008a).  However, as we show below, interactions with planetesimal
disks in systems that do not experience close encounters and scattering
events are vastly more efficient at generating resonant systems.

\subsection{MMRs in stable systems}
We now consider the generation of MMRs in stable systems that do not
experience close encounters between the planets. An empirical argument can
be made that such systems are rare amongst the progenitors of currently
observed systems, since the good agreement with the observed eccentricity
distribution requires that a large fraction of systems are unstable. At
larger radii, however, stable systems could be present and perhaps
common. In stable systems with relatively massive planets ($M_p \gtrsim
M_{\rm J}$), the planetesimal disk can act as a damping force and induce planets
to align into MMRs.  This mechanism is not efficient for low-mass planets
because the planetesimal disk has enough angular momentum to cause
excessive radial migration and drive the planets apart.  However, gentle
migration for $M_p \sim M_{\rm J}$ is very effective at forming resonant chains
reminiscent of the Galilean satellites of Jupiter (Paper 1).

\begin{figure*}
\epsscale{0.6}
\centerline{\plotone{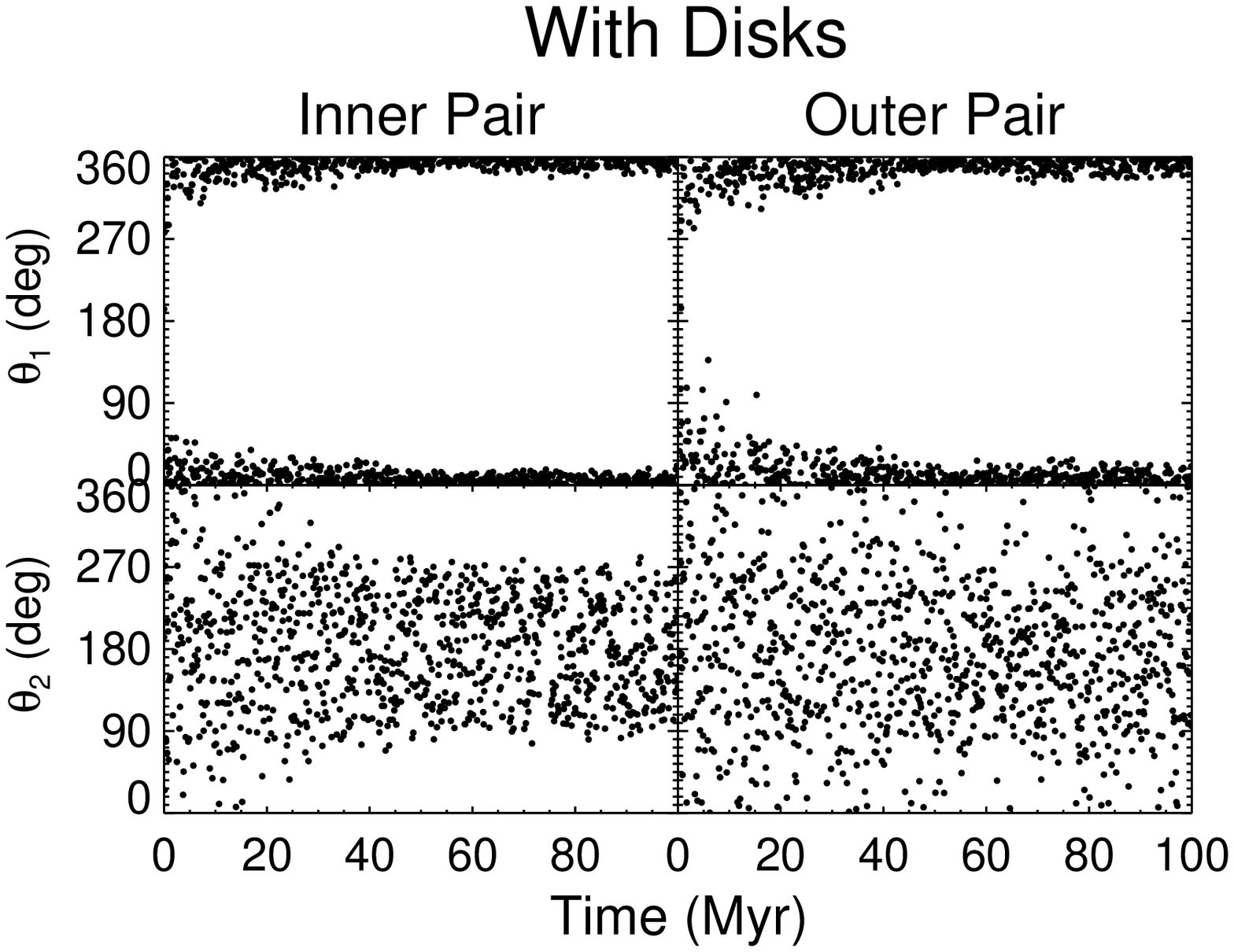}\plotone{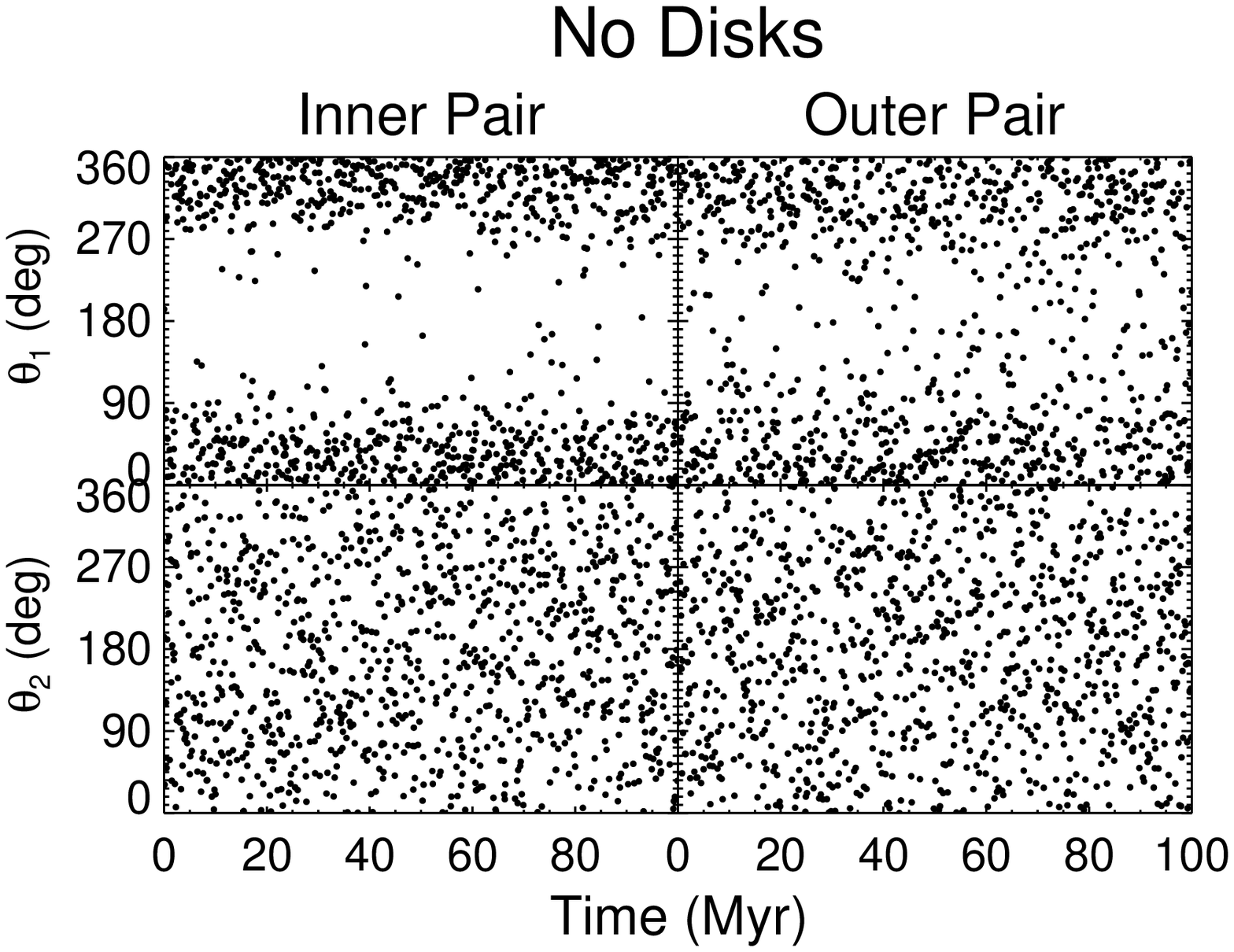}}
\caption{Example of a 4:2:1 resonant chain in a system of three planets
plus an exterior planetesimal disk.  Shown are the resonant angles
$\theta_1 = 2\,\lambda_1 - \lambda_2 - \varpi_1$ and $\theta_2 =
2\,\lambda_1 - \lambda_2 - \varpi_2$ for the inner (left) and outer (right)
pairs of planets over the 100 Myr evolution of the simulations.  The left
panel shows the system including the planetesimal disk, and the right panel
has the planets on identical starting orbits but without the planetesimal
disk.  The simulation is drawn from the {\tt Mixed1} set, and the planetary
masses are, in increasing orbital distance, 0.98, 1.42, and 1.26 $M_{\rm J}$.}
\label{fig:reschain}
\epsscale{1}
\end{figure*}

Figure~\ref{fig:reschain} (left panel) shows one example of a 4:2:1
resonant chain for a {\tt Mixed1} simulation with three $\sim$ Jupiter-mass
planets.  The $\theta_1$ resonant argument is librating with a small
amplitude for both the inner and outer pairs of planets, and for the inner
pair $\theta_2$ begins to librate after $\sim$ 20 Myr and its libration
amplitude decreases over the next 10-20 Myr.  The damping effect of the
planetesimals is clearly seen as the system moves deeper into the resonance
with time, i.e., the libration amplitude of the two resonant arguments
decreases.  The right panel of Fig.~\ref{fig:reschain} shows the same
system with no planetesimal disk.  It is clear that this system started
with both pairs of planets close to the 2:1 MMR (see
Fig.~\ref{fig:mmr-delta}) but without the disk the planets do not evolve
into the resonance.

Table 5 shows the number of resonances and resonant chains that were set up
in each set of {\em stable} simulations with planetesimal disks, and that
number as a fraction of the total number of stable simulations.  For
systems with planet masses $M_{\rm S} \lesssim M_p \lesssim 3 M_{\rm J}$, between
roughly 50 and 75 percent of all stable systems ended up with at least one
MMR!  In addition, up to forty percent of systems were in resonant chains
involving all three planets.  Recall that our definition of resonance
requires at least one resonant argument to librate with an amplitude $A
\leq 150^\circ$.  The second row for each set of simulations shows the
number and fraction of systems that were deep in resonance, with $A \leq
60^\circ$.  Roughly half of the MMRs were deep in resonance, independent of
their configuration in the system: both pairs of resonant planets were deep
in resonance for about one quarter of the resonant chains.

The vast majority of resonant systems are in the 2:1 MMR, although for
lower masses the 3:2 MMR is also common.  This is simply because of the
location of strong resonances in initial planetary separations vs. mass
(Fig.~\ref{fig:mmr-delta}).  More [less] massive planets start closer to the
2:1 [3:2] MMR, and so given the limited amount of radial migration, tend to
end up in that same resonance.  The second order MMRs were comprised almost
exclusively of the 5:3 MMRs, although those were much rarer than first
order MMRs.  Depending on the case, about half or slightly less of the
resonances were deep, meaning that the libration amplitude was $60^\circ$
or less. Systems that were {\em extremely} deep in the resonance, however, 
were rare: indeed only a single case with $A < 10^\circ$ was found in the 
entire ensemble of simulations. This is of interest since dynamical 
modeling of the GJ~876 system shows that one of the resonant angles 
librates with an amplitude that is, at most, $\theta = 7^\circ \pm 2^\circ$ 
(Laughlin \etal 2005). Taken at face value, one concludes that capture into 
resonance due to gas disk migration is not only the favored explanation 
for the origin of the GJ~876 system itself, but that gas rather than 
planetesimal effects would also be needed to explain hypothetical analogs 
of GJ~876 that might be discovered very deep in resonance at large orbital 
radii. We caution, however, that we have not demonstrated that the 
distribution of libration amplitudes is independent of numerical details 
of our simulations, and in particular that it is independent of the 
mass of particles used to represent the disk. Further simulations would be 
needed to study the predicted distribution of $A$ in more detail.

\begin{figure}
\epsscale{1.2}
\centerline{\plotone{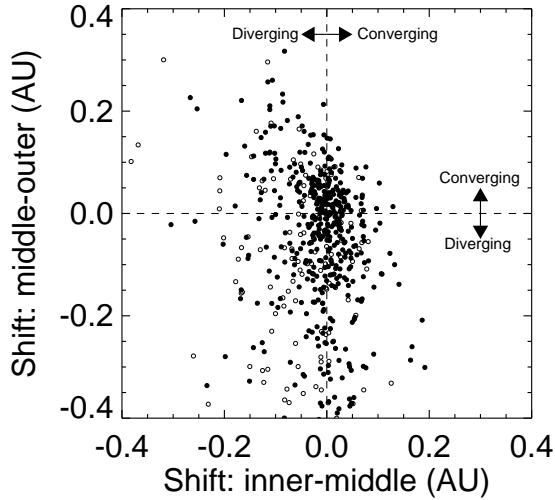}}
\caption{Semimajor axis shifts (i.e., final minus starting values) for the
planets in the stable {\tt Mixed1} and {\tt Mixed2} systems.  Positive
shift values indicate a convergence of the two planets and negative values
indicate a divergence; i.e., positive $y$ axis values indicate that the
middle and outer planets converged over the course of the simulation, and
positive $x$ axis values mean that the inner and middle planets converged.
Filled circles are systems that contain at least one pair of resonant
planets, and open circles are nonresonant systems.}
\label{fig:mmr-shift}
\epsscale{1.0}
\end{figure}

What causes the resonances in these systems?  Figure~\ref{fig:mmr-shift}
shows the semimajor axis shifts of the planets in the stable {\tt Mixed1}
and {\tt Mixed2} systems: positive values indicate convergence and negative
indicate divergence.  There is a much more limited range of parameter space
for convergence than divergence, since for massive planets the stability
limit is located in the vicinity of the 3:2 MMR.  Planet pairs were limited
to converging $\lesssim 0.1$ AU, but divergence was unlimited, at least for 
the outer pair of planets (too much divergence of the inner pair could
impinge on the outer planet).  

Table 5 shows that MMRs generally occur at a higher frequency in systems
that converge rather than diverge.  However, for most giant planet
configurations, systems for which all three planets converge are rare.  The
exception are the systems with positive mass gradients ({\tt S-J-3J} and
{\tt N-S-J}), for which almost half the stable systems converged.  This is
because they contain a very massive outer planet capable of quickly
ejecting planetesimals rather than scattering them inward, thereby
migrating slightly inward.  In contrast, systems with negative mass
gradients ({\tt 3J-J-S} and {\tt J-S-N}) favor divergence because the
innermost planet is the ejector and the outer two preferentially scatter
planetesimals inward and migrate outward.  For massive equal-mass planets
({\tt J-J-J} and {\tt 3J-3J-3J}), it is often the case that the inner two
planets converge while the outer two diverge.  This is caused simply by the
outer planet scattering planetesimals in to the middle planet.  With higher
relative velocities the middle planet can more easily eject the
planetesimals, causing it to migrate inward, away from the outer planet but
toward the inner one.

\begin{table*}
\center
\caption{Mean Motion Resonances and Resonant Chains in Stable Systems$^\star$}
\begin{tabular}{|l|cccccc|}
\hline
Set & MMRs (frac) & 2:1 & 3:2 & 2nd order & MMR chains & 4:2:1 chains \\
    & deep & deep & deep & deep & deep & deep \\
\hline
   {\tt Mixed1} & 379 (0.706) & 335 (0.624) &  58 (0.108) &  20 (0.037) & 197 (0.367) & 157 (0.292) \\
                & 190 (0.354) & 174 (0.324) &  12 (0.022) &   7 (0.013) &  50 (0.093) &  47 (0.088) \\
   {\tt Mixed2} & 162 (0.348) & 127 (0.273) &  47 (0.101) &  13 (0.028) &  54 (0.116) &  28 (0.060) \\
                &  77 (0.166) &  61 (0.131) &  16 (0.034) &   5 (0.011) &  13 (0.028) &   6 (0.013) \\
 {\tt 3J-3J-3J} & 351 (0.708) & 351 (0.708) &   0 (---) &   0 (---) &  60 (0.121) &  59 (0.119) \\
                & 122 (0.246) & 122 (0.246) &   0 (---) &   0 (---) &   3 (0.006) &   3 (0.006) \\
    {\tt J-J-J} & 217 (0.764) & 213 (0.750) &   0 (---) &   6 (0.021) & 114 (0.401) & 113 (0.398) \\
                & 103 (0.363) & 101 (0.356) &   0 (---) &   2 (0.007) &  18 (0.063) &  18 (0.063) \\
    {\tt S-S-S} &  61 (0.477) &   3 (0.023) &  60 (0.469) &   1 (0.008) &  15 (0.117) &   0 (---) \\
                &  22 (0.172) &   2 (0.016) &  22 (0.172) &   1 (0.008) &   1 (0.008) &   0 (---) \\
    {\tt N-N-N} &   0 (---) &   0 (---) &   0 (---) &   0 (---) &   0 (---) &   0 (---) \\
                &   0 (---) &   0 (---) &   0 (---) &   0 (---) &   0 (---) &   0 (---) \\
   {\tt 3J-J-S} &  92 (0.571) &  92 (0.571) &   1 (0.006) &   0 (---) &   1 (0.006) &   0 (---) \\
                &  73 (0.453) &  73 (0.453) &   0 (---) &   0 (---) &   0 (---) &   0 (---) \\
   {\tt S-J-3J} &  76 (0.927) &  76 (0.927) &  21 (0.256) &  11 (0.134) &  37 (0.451) &   5 (0.061) \\
                &  63 (0.768) &  63 (0.768) &   4 (0.049) &   3 (0.037) &   7 (0.085) &   0 (---) \\
    {\tt J-S-N} &  24 (0.118) &  24 (0.118) &   0 (---) &   0 (---) &   0 (---) &   0 (---) \\
                &  12 (0.059) &  12 (0.059) &   0 (---) &   0 (---) &   0 (---) &   0 (---) \\
    {\tt N-S-J} &  23 (0.793) &  11 (0.379) &  13 (0.448) &  11 (0.379) &   7 (0.241) &   0 (---) \\
                &   8 (0.276) &   2 (0.069) &   6 (0.207) &   5 (0.172) &   3 (0.103) &   0 (---) \\

\hline
\end{tabular}
\\ 
$^\star$Each column is the number, and in parentheses the fraction, of a
systems in a given resonance that occurred for each set of {\em stable}
simulations.  Resonances are defined to exhibit libration of at least one
resonant argument with libration angle $A < 150^\circ$.  For each set of
simulations, the second row represents the same resonance but requiring
that the system be deep in the resonance, with $A < 60^\circ$.
\end{table*}

Among the simulations with radial mass gradients, MMRs are more common in
systems with a more massive outer planet (those with positive mass
gradients, {\tt S-J-3J} and {\tt N-S-J}).  This is because the massive
outer planet confines the planetary system, and the planetesimal scattering
causes the system to converge, as a whole or for one pair of planets, much
more frequently than for a lower-mass outer planet.  Indeed, the frequency
of MMRs as a whole correlates with the fraction of systems which converge
rather than diverge (see Tables 5 and 6).  Nonetheless, many MMRs are
created in divergent systems, simply because planets were initially placed
near strong MMRs (Fig.~\ref{fig:mmr-delta}).

\begin{table*}
\center
\caption{Sources of MMRs in Stable Systems$^\star$}
\begin{tabular}{|l|ccccc|}
\hline
Set & MMRs  & conv-conv & conv-div & div-conv & div-div\\
    &       &  (frac)  &  (frac)  &  (frac)  & (frac) \\
\hline
   {\tt Mixed1} &  392 & 0.11 (0.86) & 0.24 (0.79) & 0.24 (0.77) & 0.41 (0.58) \\ 
   {\tt Mixed2} &  184 & 0.05 (0.90) & 0.05 (0.78) & 0.27 (0.58) & 0.63 (0.16) \\ 
 {\tt 3J-3J-3J} &  351 & 0.23 (0.65) & 0.35 (0.73) & 0.24 (0.69) & 0.18 (0.75) \\ 
    {\tt J-J-J} &  218 & 0.29 (0.78) & 0.38 (0.75) & 0.22 (0.75) & 0.10 (0.79) \\ 
    {\tt S-S-S} &   65 & 0.00 (---) & 0.07 (1.00) & 0.19 (0.67) & 0.73 (0.34) \\ 
    {\tt N-N-N} &    0 & 0.00 (---) & 0.00 (---) & 0.00 (---) & 1.00 (0.00) \\ 
   {\tt 3J-J-S} &   93 & 0.00 (---) & 0.07 (0.92) & 0.13 (0.71) & 0.80 (0.52) \\ 
   {\tt S-J-3J} &   84 & 0.43 (0.97) & 0.33 (0.81) & 0.22 (1.00) & 0.02 (1.00) \\ 
    {\tt J-S-N} &   24 & 0.00 (---) & 0.00 (---) & 0.00 (---) & 1.00 (0.12) \\ 
    {\tt N-S-J} &   39 & 0.45 (0.62) & 0.38 (0.82) & 0.10 (0.67) & 0.07 (1.000) \\ 
\hline
\end{tabular}
\\ 
$^\star$Each column containing 'conv' or 'div' denotes the fraction of
stable simulations in which the planets are either converging or diverging.
The first 'conv'/'div' refers to the inner pair of planets and the second
to the outer pair.  The fraction in parentheses denotes the fraction of
those cases which ended up with at least one pair of resonant planets.
\end{table*}

\section{Models for the observed distributions of extrasolar planets}
In this Section we investigate quantitatively whether the observed
distributions of extrasolar planets can be reproduced from our scattering
simulations. We focus primarily on the eccentricity distribution, since
this is the most accurately measured statistical property of extrasolar
planetary systems.  At the outset we observe that we are {\em not} claiming
that our model can explain both the radial distribution and eccentricity of
observed planets. Roughly half of the observed planets orbit at radii less
than 1~AU, which is a region of semi-major axis space that is not heavily
populated by our simulations (Fig.~\ref{fig:a-e_all}). Our systems do,
however, overlap with those planets known at radii of a few AU, and these
planets appear to share roughly the same eccentricity distribution as
planets at smaller radii (Rasio \& Ford 2008).

\subsection{The problem of the mass-eccentricity correlation}
For the {\tt Mixed1} simulations we generated planet masses by sampling the
observed planetary mass function randomly and independently, using mass
limits of $M_{\rm Sat} < M < 3 \ M_{\rm J}$. Since the eccentricity distribution
that results from these initial conditions matches the data closely
(Fig.~\ref{fig:ecum_all}) -- irrespective of whether we include disks or
not -- it might appear as if there is no further problem to solve.  This
ignores the fact that the observed exoplanet eccentricity distribution is
also mass-dependent: lower-mass planets have lower eccentricities than
higher-mass planets (Ribas \& Miralda-Escud{\'e} 2007; Ford \& Rasio 2008;
Wright \etal 2009).  The evidence for this correlation appears to be
strong.  If we divide the sample of known extra-solar planets into low-mass
($M_p \sin i < M_{\rm J}$) and high-mass ($M_p \sin i > M_{\rm J}$) bins, we find a
Kolmogorov-Smirnov probability $p$ of only 0.002 that they are drawn from
the same distribution.

\begin{figure*}
\epsscale{0.6}
\centerline{\plotone{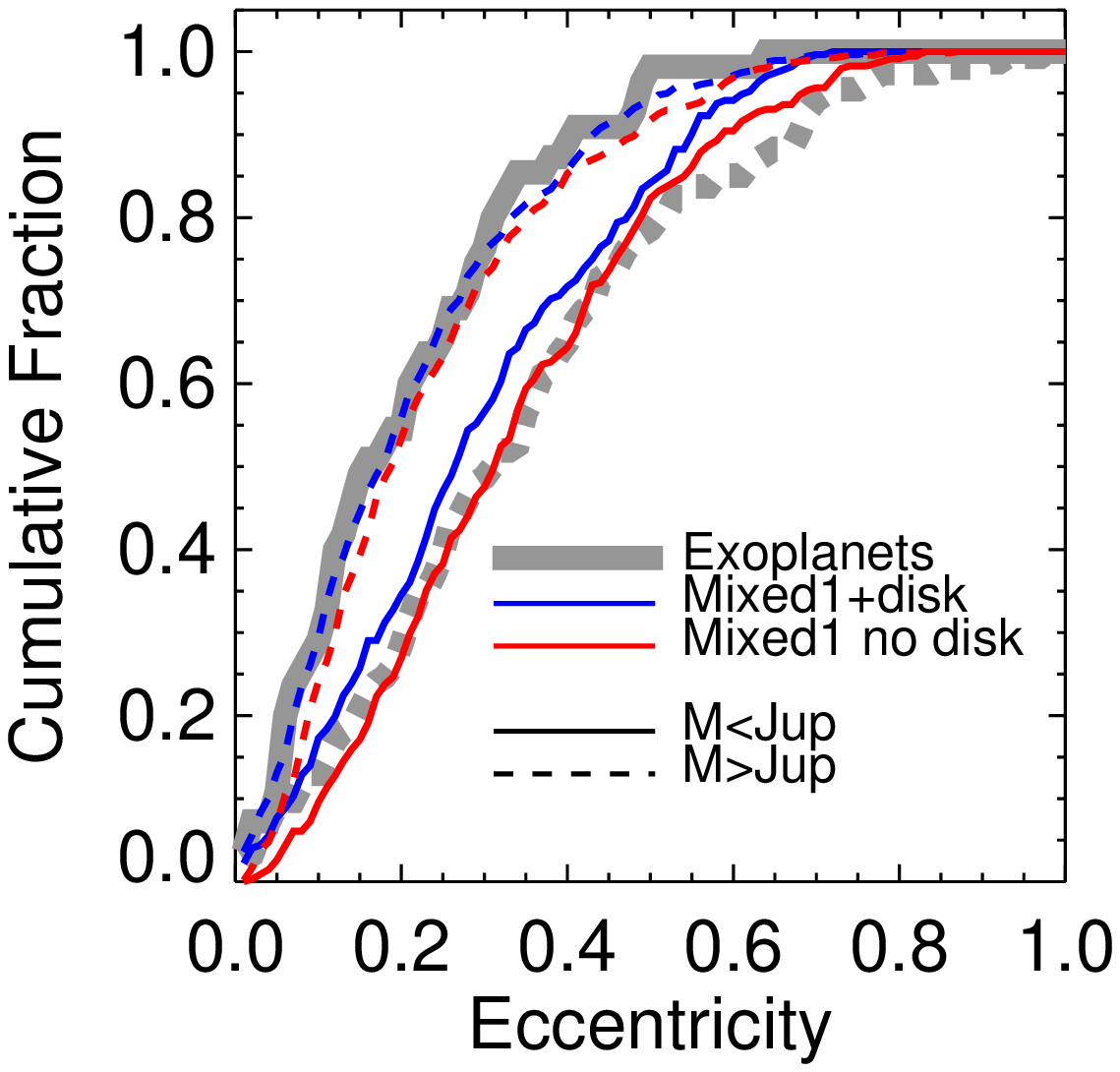}\plotone{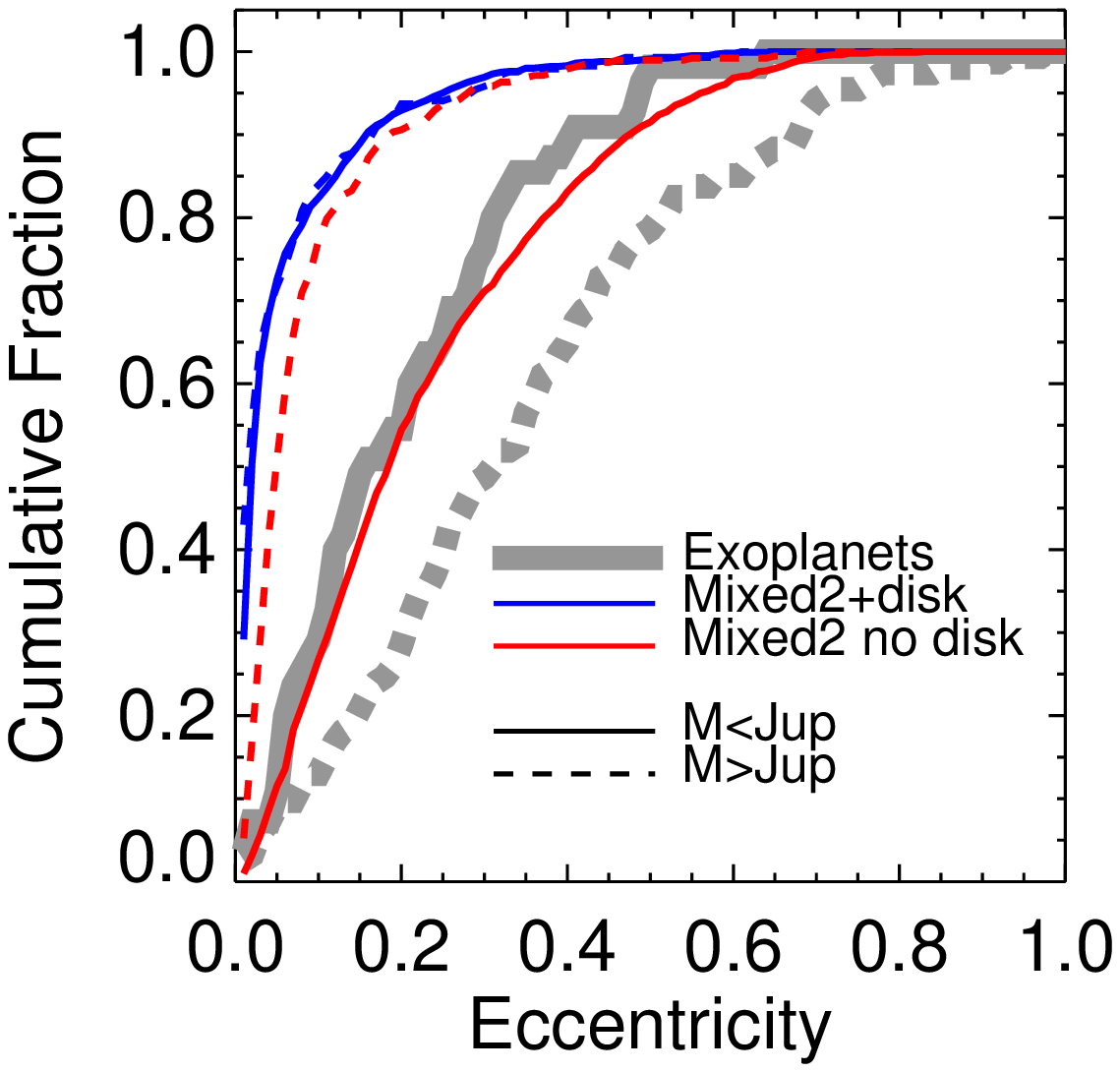}}
\caption{Cumulative eccentricity distributions for the unstable {\tt
Mixed1} (left panel) and {\tt Mixed2} (right panel), with (blue lines) and
without (red lines) disks, compared with the observed extra-solar planets
beyond 0.l AU (thick grey lines).  Each group was divided by the planet
mass: $M_p < M_{\rm J}$ are the solid lines and $M_p > M_{\rm J}$ are dashed lines. }
\label{fig:ecum_mass}
\epsscale{1.0}
\end{figure*}

Assuming that this trend is real we have a clear problem: scattering does
{\em not} predict that low-mass planets should be less eccentric than
high-mass ones. Figure~\ref{fig:ecum_mass} shows eccentricity distributions
for the {\tt Mixed1} and {\tt Mixed2} simulations divided into the same
low-mass ($M_p < M_{\rm J}$) and high-mass categories for cases with and without
disks. The predicted trend is opposite to that observed. Indeed, once the
sample has been split into mass bins it becomes clear that the good
agreement of the overall distribution is somewhat fortuitous. It occurs
because -- for the {\tt Mixed1} simulations -- the low-mass simulations match
the observed high-mass curve and vice-versa!  Changing the lower mass limit
for the mass function sampling to $10 \ M_\oplus$ only makes matters
worse. For the simulations with no disks, the more massive planets are even
less eccentric than for the {\tt Mixed1} simulations simply because the
typical mass ratio in a given scattering event is larger for the {\tt
Mixed2} simulations.  It is interesting to note that for the {\tt Mixed2}
simulations with disks the high- and low-mass planets have virtually
identical eccentricity distributions, meaning that the tendency for
low-mass planets to acquire large eccentricities during scattering is
almost exactly balanced by the eccentricity damping of the disk.

\subsection{A solution from planet mass correlations}
The above problem raises a simple question, is it possible to 
simultaneously match the observed constraints on,
\begin{itemize}
\item[(i)]
The total eccentricity distribution,
\item[(ii)]
The mass-eccentricity correlation, and
\item[(iii)]
The mass function,
\end{itemize}
using a scattering model? Although our data are not able to provide a
rigorous answer we contend that solutions {\em are} possible, either with
or without appealing to the existence of planetesimal disks. In the absence
of disks, the key assumption that we must drop is that planet masses in the
same system are chosen independently from the overall mass function. If,
instead, we allow for correlations between planet masses, the extra degree
of freedom allows us to match what is seen observationally.

For a disk-free solution, we start with the equal-mass simulations and
choose a weighted combination of the {\tt 3J-3J-3J}, {\tt J-J-J}, and {\tt
S-S-S} simulations that matches the observed mass distribution (i.e., that
contains correspondingly more low-mass than high-mass systems). By
construction this matches the mass distribution, and does a reasonable job
at matching the total eccentricity distribution (the K-S $p$ value is
0.09).  However, this combination does not match the mass-eccentricity
distribution; both the high-mass and low-mass planets have large
eccentricities, and in fact, both provide a match to the $M_p>M_{\rm J}$
exoplanets.  The high-mass planets have slightly higher eccentricities than
the low-mass planets (see Fig.~\ref{fig:ecum_all}) but the difference is
small because the highest-eccentricity systems ({\tt 3J-3J-3J}) are very
weakly-weighted to match the observed mass distribution.  Note that we
count Jupiter-mass planets in both the high-mass and low-mass bins.

We now mix in systems for which low-mass planets have lower
eccentricities than high-mass planets. In the absence of disks, this can be
done if we assume that high-mass and low-mass planetary systems are
fundamentally different. Specifically, we assume that high-mass systems form roughly
equal-mass planets yielding large eccentricities, whereas low-mass systems form
a much wider variety of systems and tend to have lower eccentricities.  To
show that this works, at least approximately, we assume that the {\tt
J-S-N} and {\tt N-S-J} simulations yield an eccentricity distribution that
is representative of the scattering among unequal-mass low-mass planets
(with masses up to, but not beyond, $M_{\rm J}$). We then add {\tt J-S-N} and
{\tt N-S-J} simulations to the total and low-mass eccentricity
distributions, but not to the high-mass distribution. We gave the {\tt
J-S-N/N-S-J} systems the same abundance as the {\tt S-S-S} systems such
that the low-mass systems are equally divided between equal-and
unequal-mass cases (except for a small contribution from the {\tt J-J-J}
simulations).

The upshot of this cookery is shown in Figure~\ref{fig:e-weight}. The
blended distribution, which is roughly consistent with the mass function,
provides a decent match to both the total eccentricity distribution and to
the mass-eccentricity distributions. The K-S test $p$ values are 0.25 (for
the total distribution), 0.09 (for the low-mass planets) and 0.20 (for the
high-mass planets). It is evident that matching the low mass planets is by
no means easy, and in fact our ``solution" is close to being statistically
unacceptable. Nonetheless, it is clearly possible to reverse the natural
tendency of scattering to generate larger $e$ for low-mass planets by
strongly correlating the masses of planets. We further believe -- though we
do not have the data to demonstrate it -- that the magnitude of the
observed mass-eccentricity correlation lies within the range that could be
accurately reproduced by fine tuning the mix of initial conditions used.

\subsection{A solution from disk damping}
\begin{figure*}
\epsscale{0.4}
\centerline{\plotone{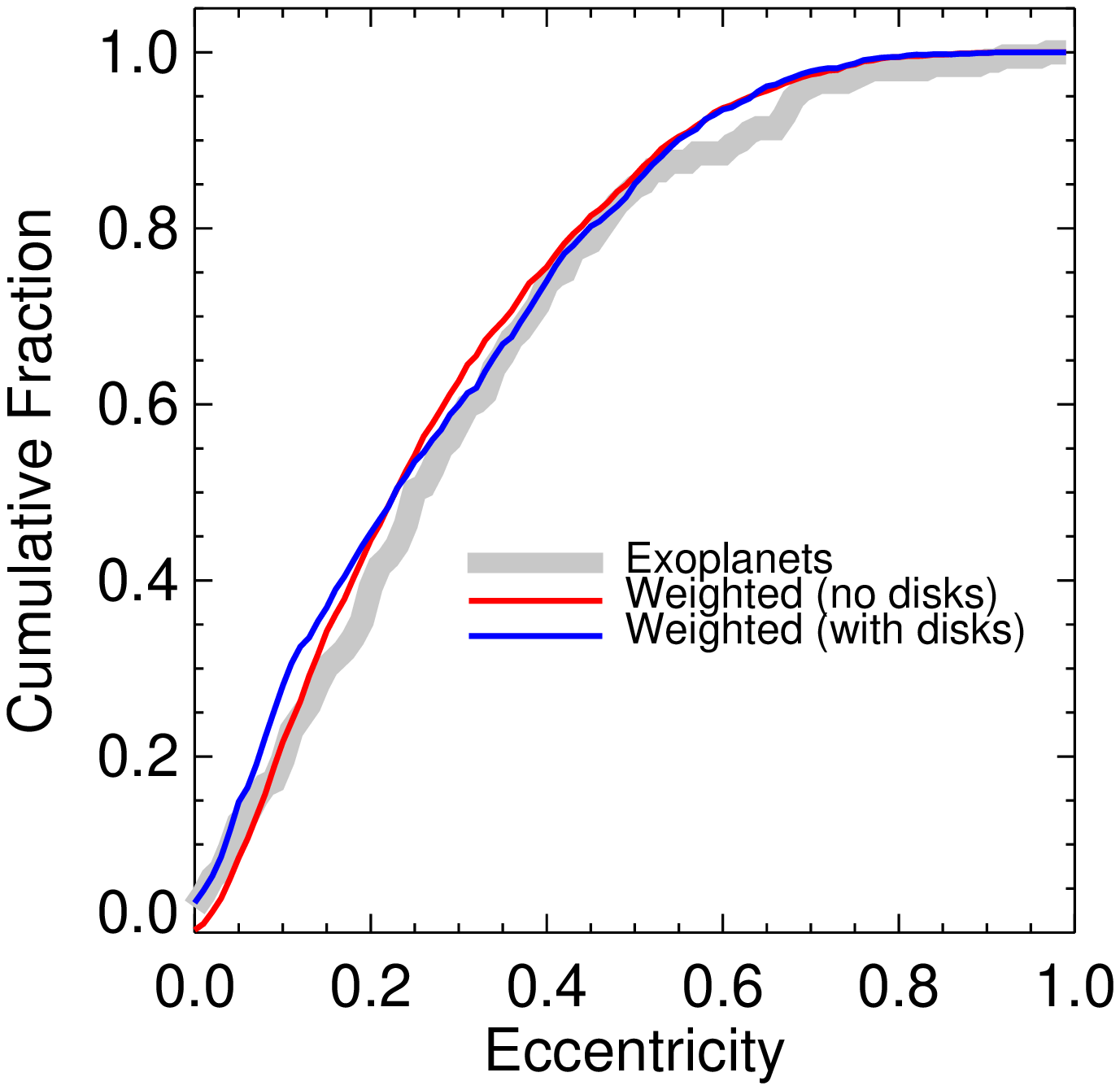}\plotone{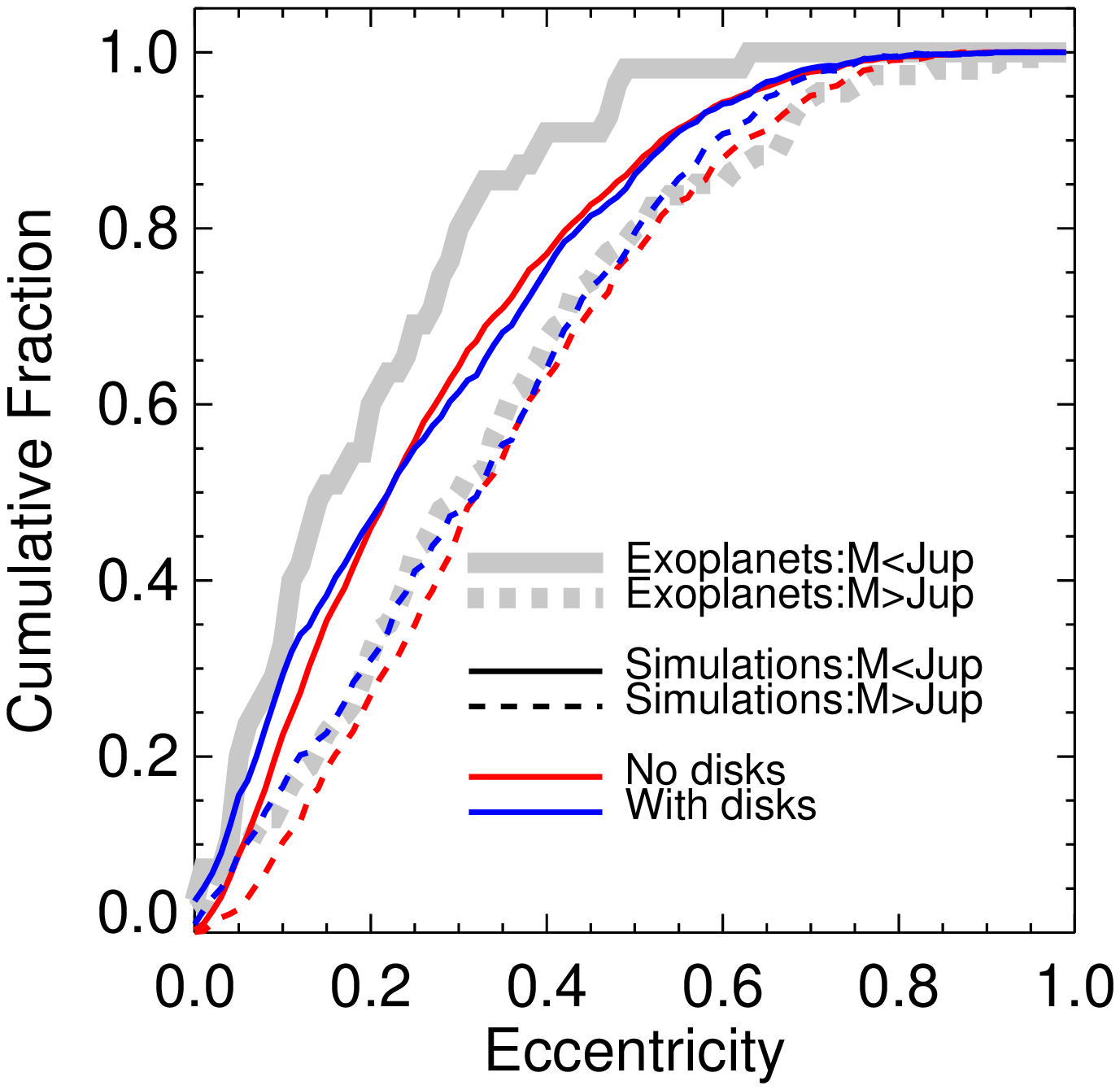}\epsscale{0.37}\plotone{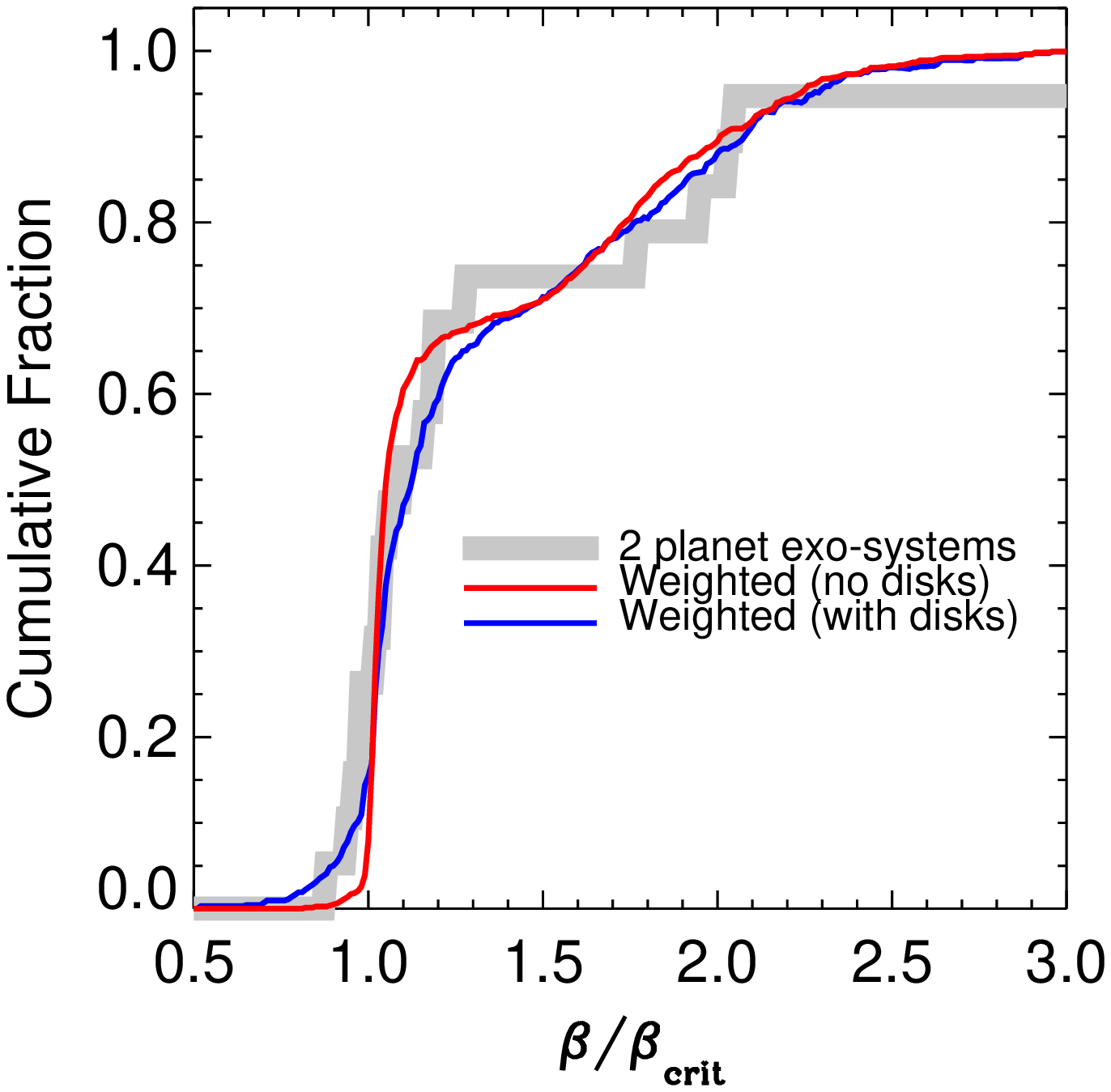}}
\caption{Comparison between our two weighted combinations of initial
conditions with the observed distributions. The left panel shows the
total cumulative eccentricity distribution, and center panel shows
the mass-eccentricity distribution: $M_p < M_{\rm J}$ are the solid lines and
$M_p > M_{\rm J}$ are dashed lines. We have only included observed planets beyond
0.l AU (thick grey lines). The right panel
shows the distributions compared with the observed $\bbc$ values of known
two planet systems.}
\label{fig:e-weight}
\epsscale{1.0}
\end{figure*}

An entirely different solution to the mass-eccentricity problem invokes the
damping effects of planetesimal disks. Disks referentially damp the
eccentricities of the smallest planets (Fig.~\ref{fig:e-m}), but, as we
have noted, the effect correlates most strongly with the total system mass
rather than that of individual planets. We therefore consider the
equal-mass cases, and (as before) blend the {\tt 3J-3J-3J}, {\tt J-J-J},
and {\tt S-S-S} simulations such as to approximate a three-bin version of
the mass function. We assume that a fraction $F=1/2$ of the systems are
affected by disks, while the rest are not. This combination provides
reasonable matches for both the total eccentricity distribution ($p$ =
0.08) as well as the low-mass ($p$ = 0.02) and high-mass ($p$ = 0.44)
planets (red curves in Fig.~\ref{fig:e-weight}). The match for the low-mass
planets can be improved by slightly increasing $F$, at the expense of the
fit to the overall distribution.

Do the two weighted distributions match other observations?  Using the same
combination of different sets of simulations we generated weighted
distributions of $\bbc$ to compare with observations, seen in the third 
panel of Fig.~\ref{fig:e-weight}.  Remarkably, both combinations of
simulations provide excellent fits to the distribution of separations in
two planet systems. The weighted case with disks was a better fit to the
$\bbc$ distribution (K-S $p$ values of 0.20 with no disks and 0.72 with
disks), largely because the cases with disks were able to fit the
distribution of values at $\bbc \leq 1$ which are dominated by resonant
systems.  Thus, the $\bbc$ distribution cannot differentiate between the
two weighted combinations of simulations we have constructed.  However, it
is interesting that both weighted distributions can match all of the
currently well-observed constraints.

\subsection{Discussion}
What can we conclude from this proof-of-principle exercise? First, 
correlations between planet masses in the same system can qualitatively 
alter some of the predictions of scattering models. Although we have 
here allowed ourselves complete freedom to impose whatever correlations 
we need, our disk-free solution in fact involves physically conceivable 
variations in the outcome of planet formation in different disks. A 
massive disk of gas and solids would be expected to form several cores 
that go on to accrete substantial envelopes, such that the outcome is 
usually a system with several very massive planets. A less massive disk, 
on the other hand, may only form one relatively low mass gas giant 
together with a number of low mass core-dominated planets. Second, 
the effects of planet-planet scattering with mass correlations, and the 
effects of planetesimals, can be almost indistinguishable, at least for the 
distributions we have considered. This should be borne in mind, 
especially as observations probe larger orbital radii where it is 
more probable that planetesimal disks play a dynamical role.

Notwithstanding the formal degeneracy of our solutions, the disk-free 
solution is favored as the origin of currently observed eccentricities. 
The reason is simply that the current sample includes many planets 
at radii that are simply too small to have experienced damping from 
plausible planetesimal disks (note that our inner disk radius of 
10~AU is already smaller than is often assumed). The match is not 
perfect but this is not surprising, since the comparison is not 
entirely fair. Both the observed and predicted eccentricities 
depend upon the Safronov number, defined as the ratio of the escape
speed from a planet's surface to the escape speed from the planetary system
at that location,
\begin{equation}
\Theta^2 = \left(\frac{M_p}{M_\star}\right) \left(\frac{a}{R}\right),  
\end{equation}
where $M_p$ and $M_\star$ are the planetary and stellar mass, $a$ is the
orbital distance and $R$ is the planetary radius (Ford \& Rasio 2008).  Our
simulations probe the region at somewhat larger $a$ than is currently
observed, such that the $\Theta$ values are correspondingly higher.  Only
at higher planetary masses are the $\Theta$ values large enough that the
variations in $\Theta$ between our simulations and the observed planets are
too small to affect the eccentricities, and it is reassuring that our
weighted fits match the high-mass planets quite well. Another point to 
bear in mind is that although the selection function of radial velocity
surveys is approximately independent of eccentricity (Cumming \etal 2008),
there are almost certainly systematic errors to add to the purely
statistical errors that we have used (e.g. Shen \& Turner 2008).

We do not expect the degeneracy between disks and mass correlations to
persist to arbitrarily large orbital radii. Disk damping becomes
increasingly more important as the semi-major axis increases, and as the
planet mass decreases. For very low mass planets the resulting reduction in
eccentricity is greater than could be explained by any plausible blend of
simulations with different conditional mass functions.  An ongoing trend to
lower eccentricities for low mass planets, as the orbital radius increases,
would thus be a strong sign of dynamical influence from planetesimal disks.

\section{Discussion and Conclusions}
We have presented an extensive study of the dynamics of planetary systems
that evolve under the joint action of planet-planet and planetesimal disk
scattering. The model is founded on the assumption that giant planets form
in marginally stable configurations interior to substantial disks of small
bodies that failed to form planets or cores. Persuasive (but
circumstantial) observational and theoretical arguments suggest generically
similar initial conditions, and physical mechanisms, describe the evolution
of typical outer planetary systems subsequent to gas disk dispersal. We
have simulated the evolution using a large set of N-body integrations,
designed such that the statistical errors on the model predictions remain
smaller than the uncertainty in observed data.  Achieving this level of
fidelity is still relatively easy for models that invoke only planet-planet
scattering, but requires substantial investment of computing resources
(amounting to several million core hours in our case) once disks are
included.

The main conclusions of our study are:
\begin{itemize}
\item[(i)]
Dynamical models for the outer Solar System such as the Nice model, in
which small body scattering dominates (Tsiganis \etal 2005), and pure
planet-planet scattering models for eccentric exoplanets, are limiting
cases of a joint model that includes both scattering processes. Future
extrasolar planet surveys are likely to uncover systems in which the two
processes were of comparable importance.
\item[(ii)]
The past dynamical effects of planetesimal disks are observable
statistically as a trend toward lower eccentricities for low mass planets
at larger orbital radius. The strongest trend is predicted to be a
correlation between eccentricity and the total mass of surviving planets in
the system. The transition radii and masses at which the effects of disks
should become apparent are not known. Our specific model predicts clear
effects for $\approx 0.5 \ M_{\rm J}$ planets at 5-10~AU. Observations in this
region of parameter space will provide important constraints on the typical
properties (the mass and inner radius) of outer planetesimal disks. Useful 
constraints would be possible from radial velocity surveys capable of 
detecting planets with radial velocity amplitudes $K \approx 5 \ {\rm m \ s}^{-1}$ 
and periods in excess of 10~years.
\item[(iii)]
Planetary inclination with respect to the initial orbital plane, and the 
mutual inclination in multiple planet systems, is 
damped by disks in the same way as is eccentricity. Our initial conditions 
(with the number of planets fixed at $N=3$) can yield significant inclinations 
when disks are unimportant ($30^\circ$ is not uncommon), but do not form 
polar or retrograde systems.
\item[(iv)]
Planet-planet scattering can reproduce the eccentricity distribution of
known extrasolar planets (Chatterjee \etal 2008; Juri\'c \& Tremaine 2008),
given only the additional assumption that the initial conditions were
unstable to close planetary encounters (this is a subset of our initial
conditions). Planetesimal disks do not destroy this agreement for planet
masses and orbital radii characteristic of the currently observed sample.
\item[(v)]
The observed variation of the eccentricity distribution with planet mass
(Ford \& Rasio 2008; Wright \etal 2009) is in the opposite sense to that
predicted by simple planet-planet scattering models. The observed trend can
probably be reproduced if we postulate strong correlations between planet
masses in the same system. Specifically, we require that one set of systems
must preferentially form massive planets of comparable masses, while the
rest form lower mass planets with a wider range of masses. This implies
that the planetary mass function is a global quantity that does not apply
on the level of single systems.
\item[(vi)]
At radii modestly larger than those currently probed by observations, the
effects of planetary mass correlations and planetesimal disks on the
eccentricity distribution can be almost identical. Additional constraints
are required to break the degeneracy and thereby determine which physical
processes are at work.
\item[(vii)]
Scattering forms dynamically packed planetary systems, in which the final
separation of two planet systems clusters close to the Hill stability
boundary (Raymond \etal 2009b). The presence of planetesimal disks causes
the final systems, on average, to be even more closely packed, at least
when only the innermost pair of surviving planets are considered.
\item[(viii)]
It is probable that the formation of planets at intermediate and large
radii (beyond 5~AU) yields an admixture of {\em stable} initial conditions
(those that do not result in close planetary encounters). Stable systems
with $\sim$Jupiter-mass planets should produce an abundance of low-order
mean motion resonances.  In many cases there should exist resonant chains
analogous to the 4:2:1 Laplace resonance among Jupiter's Galilean
satellites.  Although the observed orbital arc is short, stability analyses
suggest that the HR 8799 system (Marois \etal 2008) may be the first
example of such a resonant chain in the exoplanet population (Fabrycky \&
Murray-Clay 2008; Reidemeister \etal 2009).  We expect a large fraction of
outer, high-mass giant planets to be discovered to be in resonance,
especially if their orbits display the low eccentricities indicative of
stable systems.
\item[(ix)]
The predicted number of high mass planets on highly eccentric orbits with
large apocenter distances is unaffected by the presence of planetesimal
disks. The number of low mass planets on such orbits is, however, greatly
reduced. The existence of the high mass systems is a secure prediction of
scattering models (Veras \etal 2009; Scharf \& Menou 2009), and provides an
attractive target for direct imaging surveys of young stars.
\end{itemize}

The initial conditions and physical processes modeled in this work are 
idealized. We have assumed that the properties of the planetesimal disk 
are universal, so that variations between systems arise solely due to 
different planetary masses and architectures. To a first approximation 
this is justifiable. Giant planets can accrete vastly different envelopes -- 
from a few Earth masses in the case of ice giants up to many Jupiter masses 
at the high end -- depending upon when runaway accretion occurs relative 
to the lifetime of the gas disk (Pollack \etal 1996). The dispersion in giant planet masses is 
thus plausibly larger than the dispersion in the masses of planetesimal 
disks. In more detail, however, we expect the masses and radial extent of 
planetesimal disks to vary between systems, in a manner than is largely 
unconstrained outside the Solar System (where disk masses of $\approx 
30-50 \ M_\oplus$ are inferred; Tsiganis \etal 2005). A physical dispersion 
in disk properties would smooth the transition between the planet-planet 
and planetesimal dominated regimes that we have identified.

We have also assumed that purely N-body dynamics dominates the evolution.
Is the neglect of gas disk interactions (including migration, accretion,
and damping or excitation of eccentricity) justified? This depends upon how
the timescale for dynamical instability compares to the local gas dispersal
timescale. Using our initial conditions the onset of dynamical instability
typically requires a timescale of the order of $10^5$~yr, though there are
some systems that become unstable much more rapidly. Both observations
(Simon \& Prato 1995; Wolk \& Walter 1996) and models (Alexander, Clarke \&
Pringle 2006; Chiang \& Murray-Clay 2007) peg the global disk dispersal
timescale at $\sim 10^5$~yr, and gas is likely to disperse locally from any
particular radius substantially more rapidly. Thus, very rapid
instabilities probably do occur in the presence of residual gas (Moeckel
\etal 2008), but most take place in an almost gas-free environment.

The fact that the neglect of gas is (almost) self-consistent is, of course,
a restatement of our initial planetary separations, which were chosen
precisely to yield mostly late-onset instability. As we have noted, the
relatively widely spaced planets we start with also yield (especially in
the presence of disks) stable systems, which have quite different dynamical
properties. This causes difficulties if we try to envisage physically
motivated initial conditions that can explain the observed properties of
extrasolar planets. Matching the eccentricity distribution requires that a
large fraction of planets start in unstable configurations, but this can
only be achieved for systems that start in configurations slightly more
compact than those we have assumed. The neglect of gas may be harder to
justify in that case. Similarly, initial conditions that start with several
planets on long term unstable orbits are easiest to justify if the planets
experienced a period of resonant locking before finally detaching from
resonance. However, although there are plausible mechanisms that might
first establish and then break a resonant lock (Adams \etal 2008; Lecoanet
\etal 2009), it is hard to see why the lock should typically be broken
close to the epoch when the gas is dispersed (possibly variations in the
strength of turbulence when a weak disk becomes well-ionized may play a
role). We have no solutions to these puzzles, but they alert us to the
possibility that the typical properties of extrasolar planets at larger
orbital radii may be determined largely by a changing mixture of stable and
unstable initial conditions. As we have emphasized, the presence of stable
initial configurations results in a very high resonant fraction among
massive planets, and this is likely to be the most distinctive
observational signature that such systems exist.

Finally, we note that parallel (and possibly more powerful) constraints on
models of the type we have proposed are available via the study of dust
production and debris disks (Wyatt 2008). These aspects
can be studied by coupling our dynamical simulations to a collisional
evolution model (Booth \etal 2009). We plan to study this, along with the
influence of the evolution we have described on hypothetical terrestrial
planets, in future papers.

\section*{Acknowledgments}

We thank Google for the large amount of computing time they contributed to
this project.  We are grateful to Rory Barnes for keeping a list of current
$\bbc$ values for the community and for helping us choose which values to
use in our analysis.  We thank the anonymous referee for a careful review.  We acknowledge helpful discussions with Ruth Murray-Clay, Dimitri Veras, Geoff Marcy, Dan Fabrycky, and Brad Hansen. S.N.R. acknowledges funding from NASA Astrobiology Institutes' Virtual
Planetary Laboratory lead team, supported by NASA under Cooperative
Agreement No. NNH05ZDA001C. P.J.A. acknowledges support from the Isaac
Newton Institute for Mathematical Sciences, from the visitor program at 
Cambridge University's Institute of Astronomy, from the NSF (AST-0807471),
from NASA's Origins of Solar Systems program (NNX09AB90G), and from NASA's
Astrophysics Theory program (NNX07AH08G).  This research was supported in
part by the National Science Foundation through TeraGrid (Catlett \etal
2007) resources provided by Purdue University.

\end{document}